\documentclass{osa-article}

\journal{oe}

\articletype{Research Article}
\usepackage{subfigure}
\usepackage{lineno}
\usepackage{hyperref}

\begin{document}

\title{Release of virtual photon and phonon pairs from qubit-plasmon-phonon ultrastrong coupling system}
\author{Ting-ting Ma,\authormark{1} Yu-qiang Liu,\authormark{1} and Chang-shui Yu\authormark{1,2,*}}

\address{\authormark{1}School of Physics, Dalian University of Technology, Dalian 116024,
P.R. China\\
\authormark{2}DUT-BSU Joint Institute, Dalian University of Technology, Dalian, 116024, China}

\email{\authormark{*}ycs@dlut.edu.cn} 



\begin{abstract}
	The most important difference between ultrastrong and non-ultrastrong coupling regimes is that the ground state contains excitations. We consider a qubit-plasmon-phonon ultrastrong coupling (USC) system with a three-level atom coupled to the photon and phonon via its upper two energy levels and show that spontaneous emission of the atom from its intermediate to its ground state produces photon and phonon pairs. It is shown that the current system can produce a strong photon/phonon stream and the atom-phonon coupling plays the active role, which ensures the experimental detection. The emission spectrum and various high-order correlation functions confirm the generation of the pairs of photons and phonons. Our study has important implications for future research on virtual photon and phonon pairs creation in the ground state of the USC regime. 
\end{abstract}

\section{Introduction}
Ultrastrong coupling (USC) regime has transitioned from theoretical ideas to the experimental realization over the past few years \cite{10.1021/acs.nanolett.7b03103, doi:10.1038/nphys3906, PhysRevLett.105.237001, PhysRevB.93.214501, Nat.Photonics15, 10.1038/nphys3905} and has attracted much attention in many fields \cite{PhysRevLett.120.093602, PhysRevLett.109.193602, PhysRevResearch.4.013013, npj.QuantumInf.3.46(2017), PhysRevLett.127.073602, PhysRevA.105.023720, New.J.Phys.18.123005, PhysRevResearch.4.023016, PhysRevA.98.053834, PhysRevB.101.214414}. The tremendous difference between the USC regime and the non-USC regime is the nature of the ground state of the hybrid system \cite{PhysRevA.81.042311, PhysRevA.95.063849, PhysRevLett.110.243601, RevModPhys.91.025005, Nat.Commun.8.1715(2017), PhysRevLett.116.113601}.  For the weak and strong coupling of the light and the matter, i.e. $g/\omega_0 <0.1$, where $g$ is the coupling strength between the cavity and the atom, $\omega_0$ is the frequency of the cavity, the ground state is the empty cavity and the atom stays in the ground state. However, as the coupling strength increases into the USC regime ($g/\omega_0>0.1$), the properties of the ground state change greatly. One of the most amazing physical phenomena in nature is that the vacuum is not empty, but contains virtual particles \cite{PhysRevLett.110.243601, RevModPhys.91.025005, Adv.Quantum.Technol.3.12}. This is because, in the USC regime, the rotating wave approximation is no longer applicable \cite{PhysRevA.84.043832, PhysRevLett.109.193602}, and we have to consider the influence of the counter-rotating terms in the Hamiltonian \cite{nphys1730}. The ground state of the USC regime can be expressed as a linear superposition of bare states \cite{PhysRevA.105.053718, PhysRevD.100.125019}. Therefore, the detection and extraction of excitations in the ground state in the USC regime are important research topics \cite{RevModPhys.91.025005, PhysRevLett.110.243601, PhysRevLett.114.183601, PhysRevA.100.062501}.

The USC regime is difficult to achieve in conventional quantum-optical cavity QED but has been implemented in superconducting circuits \cite{nphys1730, PhysRevLett.105.237001, PhysRevA.96.012325}, Landau polarons \cite{Journal.of.Applied.Physics.113.136510(2013), PhysRevB.90.205309, 10.1038/nphys3850, PhysRevB.83.075309, PhysRevB.101.075301}, etc. Recently, the USC between plasmonic modes and SiO$_2$ phonons has been achieved through epsilon-near-zero (ENZ) nanocavities \cite{Nat.Photonics15}, which greatly reduces the size of the system and thus reduces the amount of material involved in realizing mid-infrared USC.  The mechanism of the SiO$_2$ phonon and the ENZ cavity coupling is different from the conventional optomechanical coupling in that the SiO$_2$ films are grown on the naked surface and sidewalls by atomic layer deposition (ALD), followed by metal cladding deposition and planarization by glancing-angle ion milling. The use of the ENZ cavity here lies in its unique advantages, the first being the control of light propagation and the improvement of the Kerr coefficient and other nonlinearities to increase the emission rate and energy transfer rate \cite{PhysRevA.96.022308}, the second being that the ENZ mode is independent of the cavity length and therefore insensitive to changes in Au film thickness \cite{Nat.Photonics15}. This new nanocavity can achieve a ratio of the coupling strength of the ENZ cavity mode with the SiO$_2$ phonon to the frequency of the SiO$_2$ vibrating phonon greater than 0.25.   The hybrid ENZ mode and vibration strong coupling mechanisms have been studied in many works \cite{acs.nanolett.8b04182,Rodrigo:21, nanoph-2020-0449, science.aau7742}. We also investigated the photon/phonon statistics of the hybrid system with the ENZ nanocavity and a trapped two-level atom \cite{PhysRevA.105.053718}. Considering the realizable USC regime in the hybrid ENZ system, it provides a new platform to detect the spontaneous release of the virtual photon/phonon pairs in the ground state.

In this paper, we consider the qubit-plasmon-phonon ultrastrong coupling system where the three-level atom is ultrastrongly coupled to the plasmon mode and SiO$_2$ phonon via its upper two energy levels (qubit). It is shown that spontaneous emission of the atom from intermediate energy levels to the ground state is accompanied by the generation of photon pairs and phonon pairs and does not require any external force. In the USC regime, the standard quantum optical master equation cannot describe the dynamic evolution of the system \cite{PhysRevLett.109.193602, PhysRevLett.110.163601, PhysRevLett.117.043601, PhysRevA.99.033809, PhysRevA.96.012325} and we have to employ the global master equation where the dissipations are characterized by the eigenoperators related to the full eigenstates of the hybrid qubit-plasmon-phonon system \cite{Adv.Quantum.Technol.3.7, PhysRevLett.126.153603, booksee.org/book/1397869, PhysRevLett.120.183601}. In addition, the conventional quantum optical correlation functions of photon/phonon fail to represent the photon/phonon detection experiments for such a hybrid system, instead one will have to use $\left\langle X^-_{c}(t)X^+_{c}(t)\right\rangle$ where $X^{+}_{c}(t)=\sum_{j,k>j}\left\langle j\right|c+c^{\dagger}| k \rangle \left|j\right\rangle \left\langle k \right|$  $(c=a,b)$ \cite{PhysRevLett.110.243601} with $a$ ($b$) and $a^{\dagger}$ ($b^{\dagger}$) denoting the annihilation operator and creation operator of the photon (phonon). Accordingly, the input-output relations of the ENZ nanocavity will also be changed \cite{PhysRevA.31.3761, PhysRevA.74.033811}.  Therefore, the virtual photon/phonon in the ground state $\left| \tilde{0} \right\rangle $ cannot be directly detected due to $ \left\langle \tilde{0} \right| X^-_{a(b)} X^+_{a(b)}\left| \tilde{0}\right\rangle=0$.  We consider the qubit-plasmon-phonon ultrastrong coupling system including a three-level atom that helps convert the virtual photon/phonon into the practically detectable photon/phonon. 
We study the photon/phonon number rate $\left\langle X_c^-(t)X_c^+(t)\right\rangle$ detected in experiments outside the ENZ nanocavity and find that the greater the coupling strength is, the greater the number of photons and phonons one can detect in the experiments. In particular, we show that the rates of photons/phonons are much larger than previously setting \cite{PhysRevLett.110.243601}, which makes it easier to detect in experiments. In addition, we also analyze the influence of different coupling mechanisms in the ENZ nanocavity on the output photon/phonon number rates \cite{Nat.Photonics15}. We find that the coupling between the atom and phonon will promote the output photon/phonon number rates, but the vibrational coupling in the ENZ nanocavity weakens them slightly. By analyzing various high-order correlation functions on photons/phonons \cite{PhysRevA.102.043701, PhysRevLett.109.193602}, we show that the photons and phonons emitted out of the ENZ cavity appear in pairs. The study offers some important insights into the photon and phonon detection in the ground state of the USC regime \cite{s41598-018-36056-1, s41467-017-01504-5, PhysRevA.89.033827}.
The remainder of this paper is organized as follows. The model and the energy levels are briefly introduced in Sec. II.  The photon/phonon number rates are mainly studied in Sec. III. The correlation of the photons/phonons is investigated by using modified correlation functions in Sec. IV. We draw our conclusion in Sec. V.
\section{Model}
\begin{figure}[tpb]
	\centering
	\includegraphics[width=0.75\columnwidth]{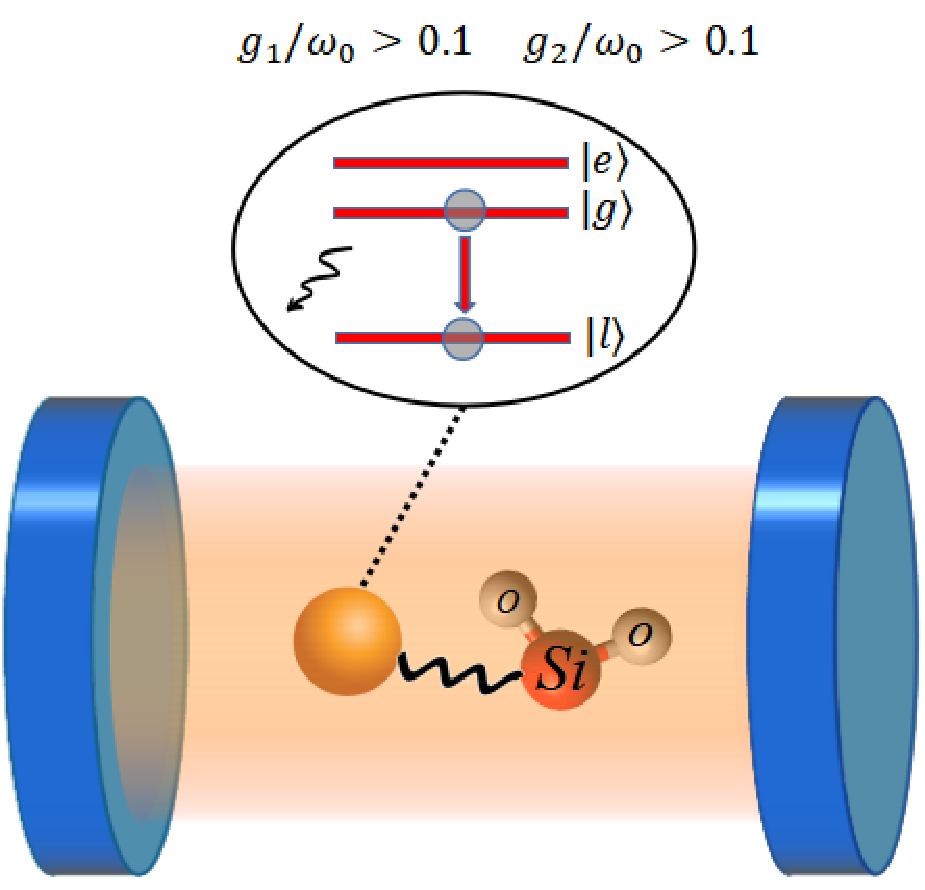}
	\caption{Schematic diagram of the hybrid qubit-plasmon-phonon system. The energy levels of the three-level atom are marked $\left\vert e \right\rangle$, $\left\vert g \right\rangle$, and $\left\vert l \right\rangle$ from high to low. The first and second excited states of the atom are coupled to the plasmon mode and SiO$_2$ phonon. Spontaneous decay process $\left| g \right\rangle \to \left| l \right\rangle$ is accompanied by the generation of photon and phonon pairs.}
	\label{model}
\end{figure}

\begin{figure}[tpb]
	\centering
	\includegraphics[width=0.8\columnwidth]{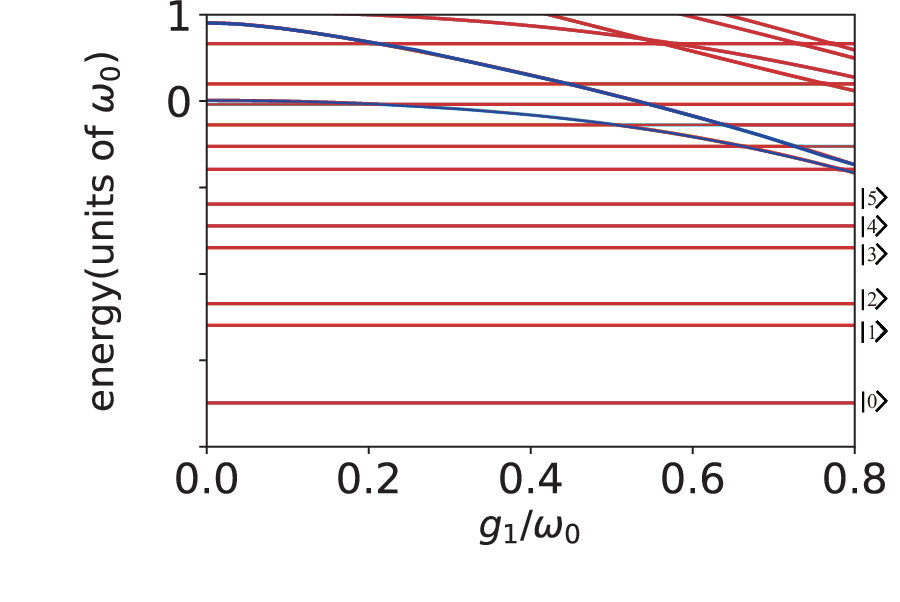}
	\caption{Lower energy levels of the full hybrid qubit-plasmon-phonon system as a function of the coupling strength $g_1$. Y-axis represents the eigenvalues of the $H$. The bending lines correspond to the eigenvalues of $H_1$ and the horizontal lines illustrate the eigenvalues of $H_l$.  Here we take $g_1=g_2$ for simplicity. $\left|0\right\rangle$ to $\left|5\right\rangle$ denote the six lowest energy eigenstates of the full hybrid system, respectively.}
	\label{spec}
\end{figure}

The model of interest sketched in Fig. \ref{model} is a hybrid qubit-plasmon-phonon system that includes a three-level artificial cascade atom interacting with an ENZ nanocavity with vibrational ultrastrong coupling between the plasmon mode and SiO$_2$ phonon. The atom is a qubit based on phase-biased flux. Two excited states and one ground state of the three-level atom  are  labeled by $\left|e\right\rangle$, $\left|g \right\rangle$, and $\left|l\right\rangle$, respectively. It should be noted that the operating frequency of our system is the mid-infrared frequency \cite{Nat.Photonics15}. The free Hamiltonian of the cavity and the atom reads
\begin{equation}
	H_0=\omega_0 a^{\dagger}a+\omega_b b^{\dagger}b +\sum_{\alpha=e,g,l} \omega_{\alpha} \sigma_{\alpha \alpha},
\end{equation}
where $\omega_0$ and $\omega_b$ are the frequency of the plasmon mode and SiO$_2$ phonon, $\omega_{\alpha}$ represents the different bare frequency of the atom and $\sigma_{\alpha \alpha}$, $\alpha=e,g,l$, denotes the projector $\left|\alpha \right \rangle \left \langle \alpha \right|$.
In the ENZ cavity, the plasmon modes are ultrastrongly coupled to SiO$_2$ phonons. In addition, we let the above energy levels $\left|e\right\rangle$ and $\left|g\right\rangle$ resembling a two-level system resonantly interact with both the plasmon mode and the SiO$_2$ phonons.  So the interaction Hamiltonian can be given by \cite{Nat.Photonics15}
\begin{align}
	&H_I=ig_C(a+a^{\dagger})(b-b^{\dagger})+g_D(a+a^{\dagger})^2\nonumber\\
	&+g_1(a+a^{\dagger})(\sigma_{eg}+\sigma_{ge})+g_2(b+b^{\dagger})(\sigma_{eg}+\sigma_{ge}),	
	\label{Ham i}
\end{align}
where $\sigma_{eg}$ represents the transition from $\left\vert g \right\rangle$ to $\left\vert e \right\rangle$ and is an ascending operator, corresponding to $\sigma_{ge}$  is a descending operator, $g_1$ is the coupling strength between the plasmon mode and the above energy of the atom \cite{PhysRevLett.109.193602, PhysRevLett.110.243601}, $g_2$ is the coupling strength between the phonon mode and the above energy of the atom \cite{10.1038/nature11821, s41586-019-0960-6},     
\begin{align}
	g_C=\frac{\omega_{p}}{2}\sqrt{\frac{\omega_{b}}{\omega_{0}}}, g_D=\frac{\omega_{p}^2}{4\omega_{0}},
\end{align}
with $\omega_p$ being the vibrational coupling constant, and $g_C$ and $g_D$ denoting the coupling constants of the plasmon and the phonon mode \cite{PhysRevB.72.115303, Nat.Photonics15}.                                                                                                                                                                                                         
One can note that the artificial atom won't interact with the ENZ nanocavity if the atom is in the ground state $\left|l\right\rangle$, thus the total Hamiltonian
$H=H_0+H_I$ can be written as two parts as $H=H_{1}+H_l$, where $H_l=\omega_l\sigma_{ll}$ and $H_{1}=H-H_l$ represents the interaction between the ENZ nanocavity with ‘two-level atom’.  As a result, the eigenstates of the system can also be divided into two parts: one is the noninteracting states $\left| n_a,n_b,l\right\rangle$, where $n_a$, $n_b$ denote the photon and phonon number, respectively, the other is the eigenstates of $H_1$ denoted by $\left| \tilde{j} \right\rangle $. The corresponding energy spectrum is intuitively illustrated in Fig. \ref{spec}. The ground state of $H_1$ is given by  $\left| \tilde{0} \right\rangle $, which is not a vacuum state \cite{PhysRevLett.116.113601, PhysRevLett.110.243601, PhysRevA.104.023109, PhysRevA.100.063827} in the USC regime. It especially can be expressed by the linear  combination of bare photon, phonon, and atom states as $\left| \tilde{0} \right\rangle =\sum_{n_an_b}{C}_{n_an_b g}\left|n_a, n_b, g\right\rangle+\sum_{\tilde{n}_a\tilde{n}_b}\tilde{C}_{\tilde{n}_a\tilde{n}_b e}\left|\tilde{n}_a, \tilde{n}_b, e\right\rangle$. A different coupling constant will change the Hamiltonian as well as the ground state  $\left| \tilde{0} \right\rangle $. 

\begin{figure}[tpb]
	\centering
	\subfigure[]{\includegraphics[width=0.49\columnwidth]{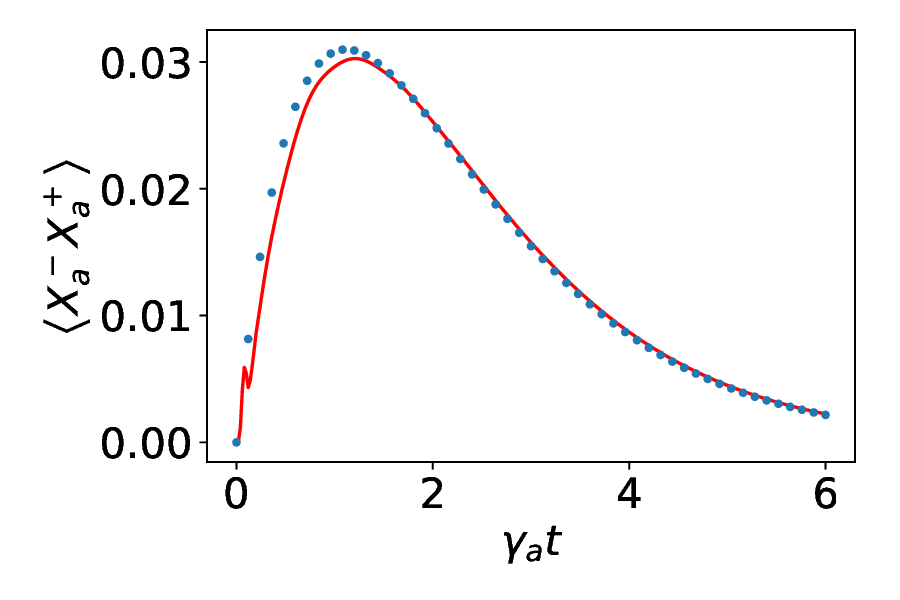}}
	\subfigure[]{\includegraphics[width=0.49\columnwidth]{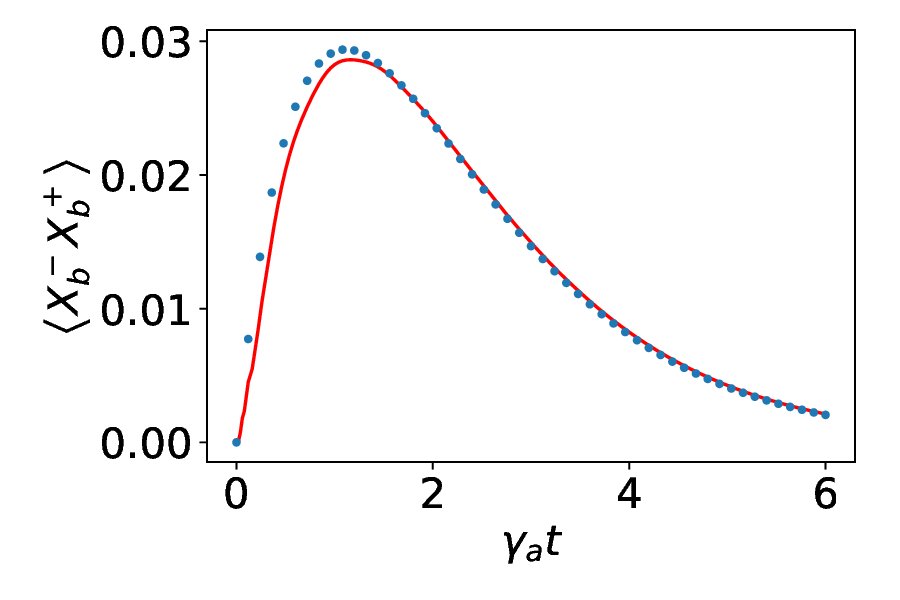}}
	\caption{Comparison of the evolutions from the $\pi$-pulse prepared initial state (red line) and from the initial state $\left| \tilde{0} \right\rangle$ (blue dots). (a) represents the time evolution of the mean photon number $\left\langle X_a^{-}(t)X_a^{+}(t)\right\rangle$. (b) represents the time evolution of the mean phonon number $\left\langle X_b^{-}(t)X_b^{+}(t)\right\rangle$. The internal dissipations of the system are taken as $\gamma_a=\gamma_b=\gamma_{eg}=\gamma_{gl}=0.02\omega_0$, $\omega_{gl}=3.5\omega_0$, $\omega_p=0.25\omega_0$, $g_1=g_2=0.6\omega_0$.}
	\label{Comp}
\end{figure}

Here we consider that the system starts with the lowest state $\left| \tilde{0} \right\rangle$ of the Hamiltonian $H_1$, which is close to the bare state $\left|0,0,g\right\rangle$. This state can be prepared by exciting the atom with $\pi$ pulse and the driving Hamiltonian is given by $H_p=\frac{\sqrt{3\pi}}{\sigma} e^{-t^2/(2\sigma^2)}\cos\omega t (\sigma_{gl}+\sigma_{lg})$ with $\sigma=5/\omega_0$ and $\omega=\omega_s$, which is marked in Fig. \ref{specc}. An intuitive comparison of the evolutions of $\left\langle X_c^{-}(t)X_c^{+}(t)\right\rangle(c=a,b)$ from the $\pi$-pulse prepared initial state and from the initial state $\left| \tilde{0} \right\rangle$  is plotted in Fig. \ref{Comp}.

\section{Creation of photon and phonon pairs}
In the USC regime, various components of the system ultrastrongly interact with each other and are inseparable \cite{NOH2019350}. To consider the dynamics of the system which includes a variety of dissipation channels, one will have to employ the global master equation \cite{PhysRevLett.110.243601, PhysRevA.105.053718, AIP.Advances.10.025106, ref2004Gemma}
\begin{equation}
	\dot\rho(t) = i [\rho(t), H] + \sum_c  \mathcal{L}_{c}\rho(t),
	\label{master}
\end{equation}
where $ \mathcal{L}_{c}\rho(t) = \sum_{j,k>j}\Gamma^{j k}_{c} \mathcal{D}[|j \rangle \langle k|]\rho(t)$ represents the superoperators that describe losses via various dissipation channels with the dissipator $\mathcal{D}[\mathcal{O}]\rho = \frac{1}{2} (2 \mathcal{O}\rho\mathcal{O}^{\dagger}-\rho \mathcal{O}^{\dagger} \mathcal{O} - \mathcal{O}^{\dagger} \mathcal{O}\rho)$. Note that $c$ takes different forms: $c=a$ represents the plasmon mode, $c=b$ represents the phonon mode, and $c=ge, gl$ represent the three-level artificial atom. The relaxation coefficients $\Gamma^{j k}_{c} = 2\pi d_{c}(\Delta_{k j}) \alpha^{2}_{c}(\Delta_{k j})| C^{(c))}_{j k}|^2$ can be simplified to $ \Gamma^{j k}_{c} = \gamma_{c}  |C^{(c)}_{j k}|^2$ with $C^{(c)}_{j k} =  \langle j |(c + c^{\dagger})| k \rangle$. It should be emphasized here that the master equation represented by Eq. \ref{master} is used under the condition that the system is ultrastrong coupled within the system, and the rotating wave approximation no longer applies. The system is weakly coupled to the environment, so the Born-Markov approximation can still be applied \cite{PhysRevA.84.043832}.

\begin{figure}[tpb]
	\centering
	\subfigure[]{\includegraphics[width=0.49\columnwidth]{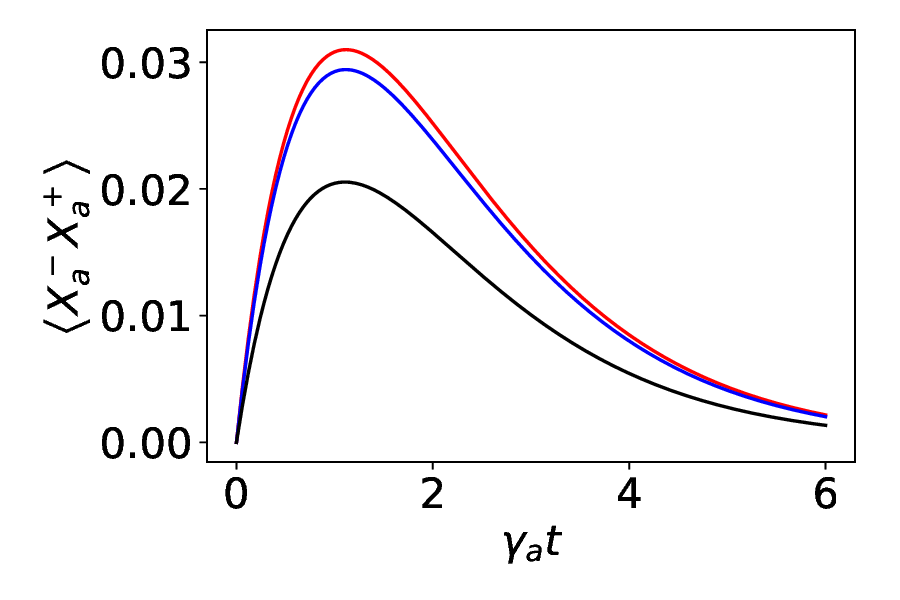}}
	\subfigure[]{\includegraphics[width=0.49\columnwidth]{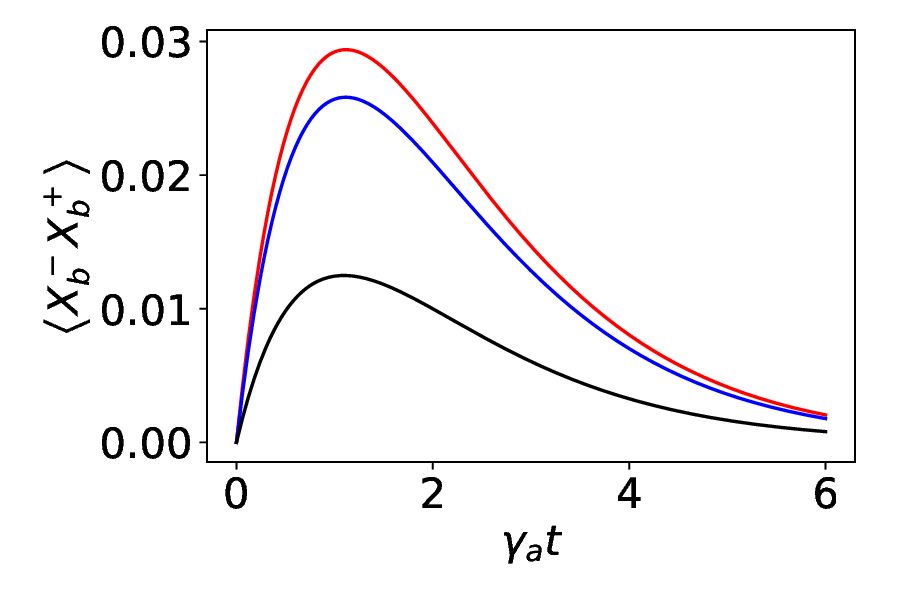}}
	\subfigure[]{\includegraphics[width=0.49\columnwidth]{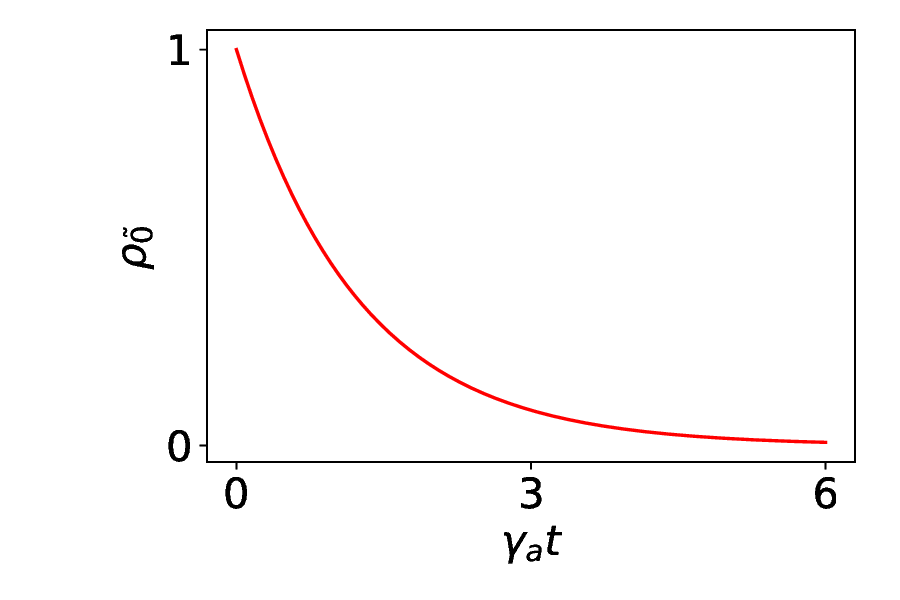}}
	\subfigure[]{\includegraphics[width=0.49\columnwidth]{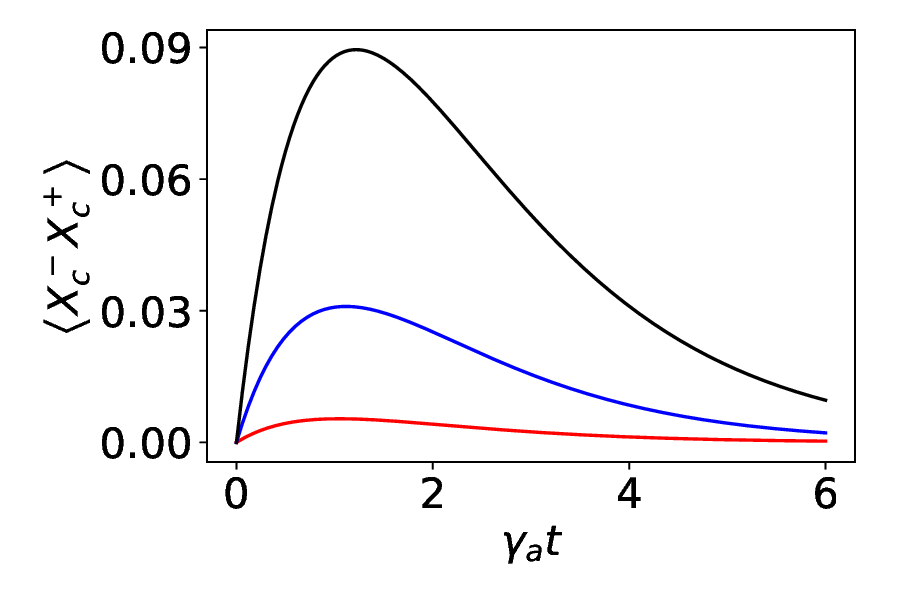}}
	\subfigure[]{\includegraphics[width=0.49\columnwidth]{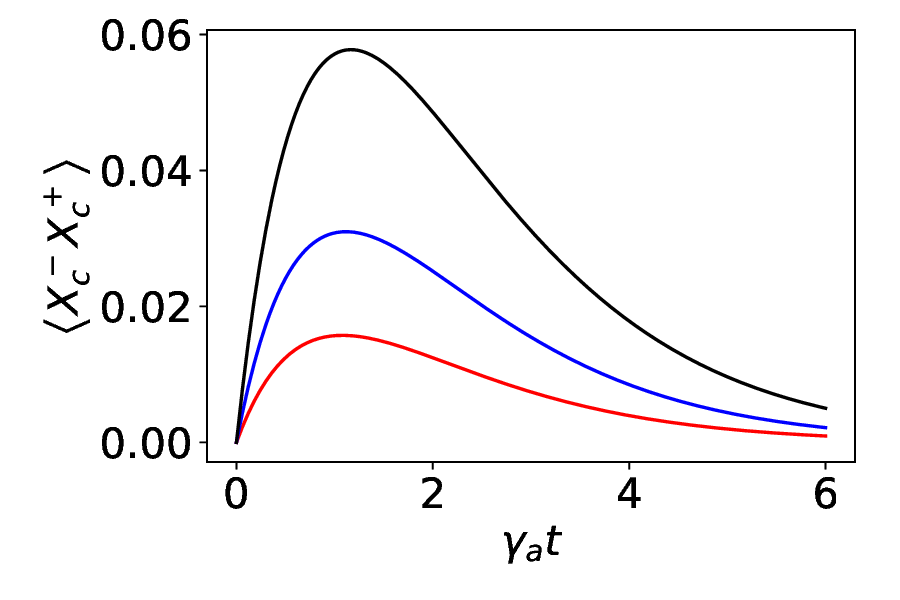}}
	\subfigure[]{\includegraphics[width=0.49\columnwidth]{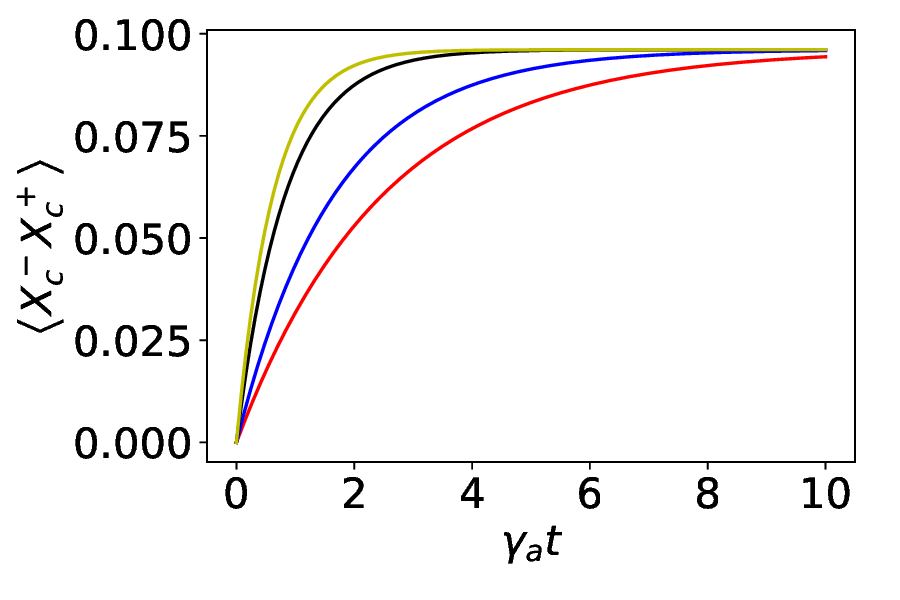}}
	\caption{The evolution of the intra-cavity mean photon and phonon numbers $\left\langle X_a^{-}(t)X_a^{+}(t)\right\rangle$  (a) and $\left\langle X_b^{-}(t)X_b^{+}(t)\right\rangle$ (b) for different vibrational coupling constants $\omega_p=0.25\omega_0$ (red line), $\omega_p=0.4\omega_0$ (blue line), $\omega_p=0.8\omega_0$ (black line), $\gamma_a=\gamma_b=\gamma_{eg}=\gamma_{gl}=0.02\omega_0$, $g_1=g_2=0.6\omega_0$.  (c) The population dynamics of the initial state $\left \vert \tilde{0} \right\rangle$. (d) The evolution of the intra-cavity mean photon and phonon numbers $\left\langle X_c^{-}(t)X_c^{+}(t)\right\rangle$, $(c=a,b)$  for different coupling strengths $g_1/\omega_0 = 0.4$(red line), $0.6$(blue line), $0.8$(black line), with $g_1=g_2$, $\omega_p=0.25\omega_0$. (e) $\left\langle X_c^{-}(t)X_c^{+}(t)\right\rangle$, $(c=a,b)$ versus $\gamma t$ corresponds to different atom-phonon coupling $g_2=0.4\omega_0$ (red line), $g_2=0.6\omega_0$ (blue line), $g_2=0.8\omega_0$ (black line), where $g_1=0.6\omega_0$, $\omega_p=0.25\omega_0$. (f) $\left\langle X_c^{-}(t)X_c^{+}(t)\right\rangle$, $(c=a,b)$ versus $\gamma t$ for different spontaneous emission rates $\gamma_{gl}=0.01\omega_0$ (red line), $\gamma_{gl}=0.015\omega_0$ (blue line), $\gamma_{gl}=0.03\omega_0$ (black line) and $\gamma_{gl}=0.04\omega_0$ (yellow line) with $g_1=g_2=0.6\omega_0$, $\gamma_a=\gamma_b=0$, $\gamma_{eg}=0.02\omega_0$, $\omega_p=0.25\omega_0$.}
	\label{photon}
\end{figure}

To detect the emitted photons and phonons from the ENZ nanocavity in experiments, one will have to study the corresponding average emitted particle number $\left\langle c_{out}^{\dagger}c_{out}\right\rangle (c=a,b)$, which are proportional to $\left\langle X_c^{-}(t)X_c^{+}(t)\right\rangle(c=a,b)$ if the input is in the vacuum state, based on the input-output relations $c_{out}(t)=c_{in}(t)-\sqrt{\gamma_c}X_c^{+}(t) (c=a,b)$ in the USC regime \cite{PhysRevA.74.033811, Horoshko:98}.  One can note that the coupling strength of the plasmon mode and SiO$_2$ phonon mode  $g_C$, and the two-photon coupling strength $g_D$  depend on the vibrational coupling constant $\omega_p$. It is easy to show that $g_C$ is greater than $g_D$, which means the coupling effect between the plasmon mode and the phonon is stronger in these two coupling mechanisms. In this sense, the difference between the dynamics of photon and phonon could be quite subtle for the weak vibrational coupling  $\omega_p$, if other parameters have no obvious difference.  In Fig. \ref{photon} (a) and (b), we show the number rates of output photons  $\left\langle X_a^{-}(t)X_a^{+}(t)\right\rangle$ and phonons $\left\langle X_b^{-}(t)X_b^{+}(t)\right\rangle$ corresponding to different vibrational coupling constants $\omega_p=0.25\omega_0, 0.4\omega_0, 0.8\omega_0$, respectively. All subsequent calculations are in the case of resonance. Other parameters are taken as $\gamma_a=\gamma_b=\gamma_{eg}=\gamma_{gl}=0.02\omega_0$, $\omega_{gl}=3.5\omega_0$. The initial state is the lowest state $\left| \tilde{0} \right\rangle$. It is also shown that with the increase of the vibrational coupling constant, the maximum values of the output photon and phonon number rates will decrease slightly, which shows the negative effect of the vibrational coupling and is a noticeable phenomenon.
In addition, comparing Fig. \ref{photon} (a) and (b) shows that for $\omega_p=0.25\omega_0$, the evolutions of photon and phonon are almost the same, and with $\omega_p$ increasing, their difference increases, which coincides exactly with our analysis.  Ref. \cite{Nat.Photonics15} shows that the plasmon-phonon coupling in the ENZ nanocavity can be realized in the ultrastrong regime in experiments, which corresponds to the vibrational coupling constant ($\omega_p$) of this nanocavity around $0.25\omega_0$. However, only when the vibrational coupling constant $\omega_p$ increases to such an extent that the coupling between the photon and the photon associated with $g_D$ also enters into an ultrastrong coupling regime or even deep coupling regime that the results for the photon and phonon become significantly different, which could be difficult to achieve in current experiments.  Therefore, in the following, we will only take the vibrational coupling constant $\omega_p$ to be $0.25\omega_0$ if not specified. That is, $\left\langle X_c^{-}(t)X_c^{+}(t)\right\rangle$ in the latter figures can be roughly considered to simultaneously represent photon and phonon numbers.

The output  photon and phonon beam comes from spontaneous decay of the state  $\left| \tilde{0} \right\rangle$. Taking into account dissipations, we have studied the population evolution of the $\left \vert \tilde{0} \right\rangle$ in Fig. \ref{photon} (c), which indicates the decay of the population. However, one can see that the output phonon and photon beams reach a maximum value before the exponential decay to the tail by comparing Fig. \ref{photon} (c) with Fig. \ref{photon} (a) and (b). This allows virtual photons and phonons in the  $\left| \tilde{0} \right\rangle $ to be well detected.   Since the three-level atom plays the key role in the detection of emitted photons/phonons,  we also studied the evolution of $\left\langle X_c^{-}(t)X_c^{+}(t)\right\rangle $ in Fig. \ref{photon} (d), where we take the couplings of the photon and phonon to the atom to be equal. The red, blue, and black lines represent different coupling strengths $g_1=g_2$ between photon/phonon and the atom $0.4\omega_0$, $0.6\omega_0$,  and $0.8\omega_0$, respectively.  Fig. \ref{photon}(d) indicates that as the coupling strength increases, the maximum number of the detected photons and phonons increases. When the coupling strength $g_1=0.8\omega_0$, the maximum value of $\left\langle X_c^{-}(t)X_c^{+}(t)\right\rangle$ can reach $0.09$, which is obviously larger than that in the previous scheme \cite{PhysRevLett.110.243601}. So the current technology should be easy to experimentally detect the photons and phonons using quadrature amplitude detectors \cite{PhysRevLett.105.133601, 10.1038/nphys1845, PhysRevLett.105.100401}. The coupling of the atom and the SiO$_2$ phonon mode is another important feature of the current system. The coupling of the atom and the phonon has also a significant influence on the released photons and phonons from the ENZ cavity, which is illustrated in Fig. \ref{photon} (e), where the coupling of the atom and the photon is taken as $g_1=0.6\omega_0$ and the coupling strength between phonon and atom is taken as $g_2=0.4\omega_0, 0.6\omega_0, 0.8\omega_0$ corresponding to red, blue and black lines, respectively. Apparently, the output phonon and photon number rates increase with the increase of the coupling between the phonon and atom, which indicates that the phonon-atom coupling promotes the experimentally detectable phonon and photon number rates.
We also investigate the atomic spontaneous decay on the mean photon (phonon) number $\left\langle X_{c}^{-}(t)X_{c}^{+}(t)\right\rangle$, which is plotted in Fig. \ref{photon} (f). To illustrate the effect of atomic decay well, we let $\gamma_a=\gamma_b=0$. It can be found that the maximum value of the mean photon and phonon numbers does not hinge on $\gamma_{gl}$ in the absence of the photon and phonon losses. $\gamma_{gl}$ only affects the speed of increase, the larger its value is, the faster $\left\langle X_c^{-}(t)X_c^{+}(t)\right\rangle$ rises.

\begin{figure}[tpb]
	\centering
	\subfigure[]{\includegraphics[width=0.49\columnwidth]{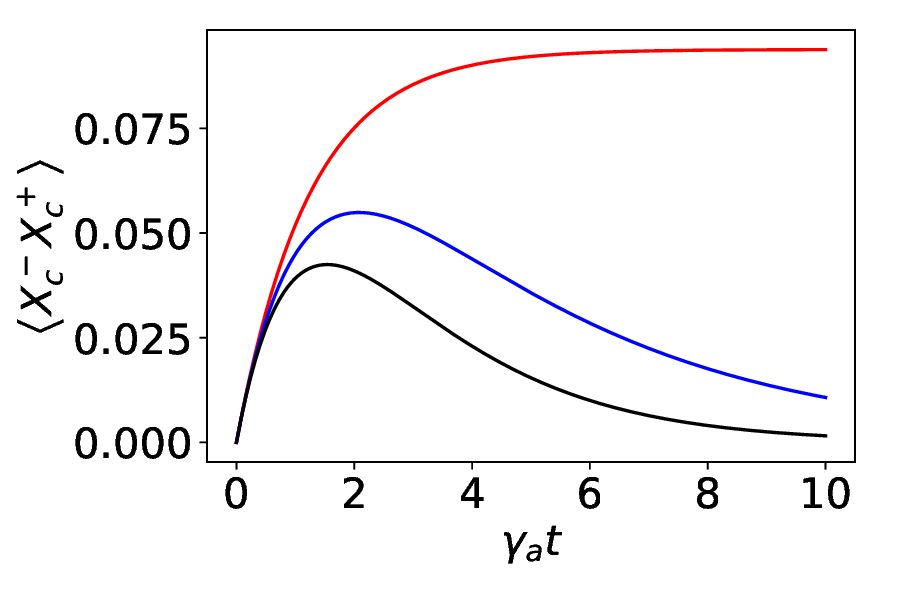}}
	\subfigure[]{\includegraphics[width=0.49\columnwidth]{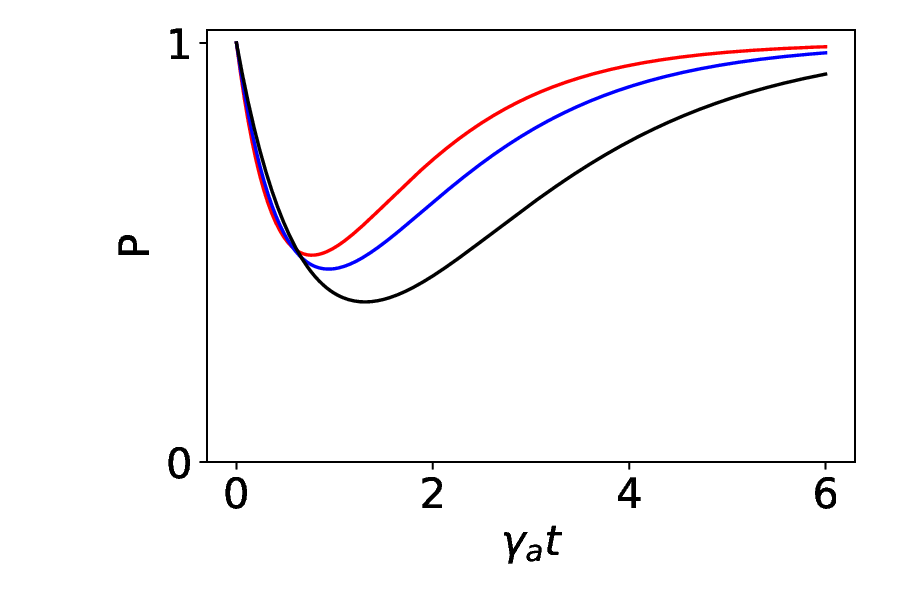}}
	\caption{(a)$\left\langle X_c^{-}(t)X_c^{+}(t)\right\rangle$, $(c=a,b)$ versus $\gamma t$ for different the loss of the cavity $\gamma_{c}=0$ (red line), $\gamma_{c}=0.01\omega_0$ (blue line), and $\gamma_{c}=0.02\omega_0$ (black line). (b) Purity versus time. The black, blue, and red lines correspond to the case of Fig. 4(d) in the main text, respectively.}
	\label{blue}
\end{figure}

Meanwhile, for the completeness of the analysis, we also investigated the effect of the loss of the optical cavity, and the results are shown in Fig. \ref{blue}. When optical cavity and vibration mode losses are not considered, the average photon number and phonon number remain constant after reaching the maximum value, and when optical cavity and vibrational mode losses are considered, the maximum values of the average photon number and phonon number decrease, and the larger the loss, the smaller the maximum value, and decay rapidly after reaching the maximum values.

To further confirm the validity of our scheme for releasing photon-phonon pairs, we add the analysis of the purity, which is defined as:
\begin{align}
	P=\mathrm{Tr} [\rho^2]
\end{align}

Purity is used to measure the degree of mixing of quantum states \cite{PhysRevA.98.053811}. The lower the purity, the greater the mixing of the states, the greater the coupling effect achieved between the system and the environment. As shown in Fig. \ref{blue} (b) that the point where the maximum number of photon-phonon pairs is released corresponds to the lowest point of purity, which means that the mixing between the system and the environment is maximized when the maximum number of photon-phonon pairs are released. Comparing Fig. \ref{photon} (d) and Fig. \ref{blue} (b), we can see that when the coupling strength is relatively small and the released photon-phonon pairs are relatively small, the purity of the photon-phonon pairs increases accordingly, which illustrates the validity of our scheme.

\section{The correlation of the photon and phonon pairs}

\begin{figure}[tpb]
	\centering
	\subfigure[]{\includegraphics[width=0.49\columnwidth]{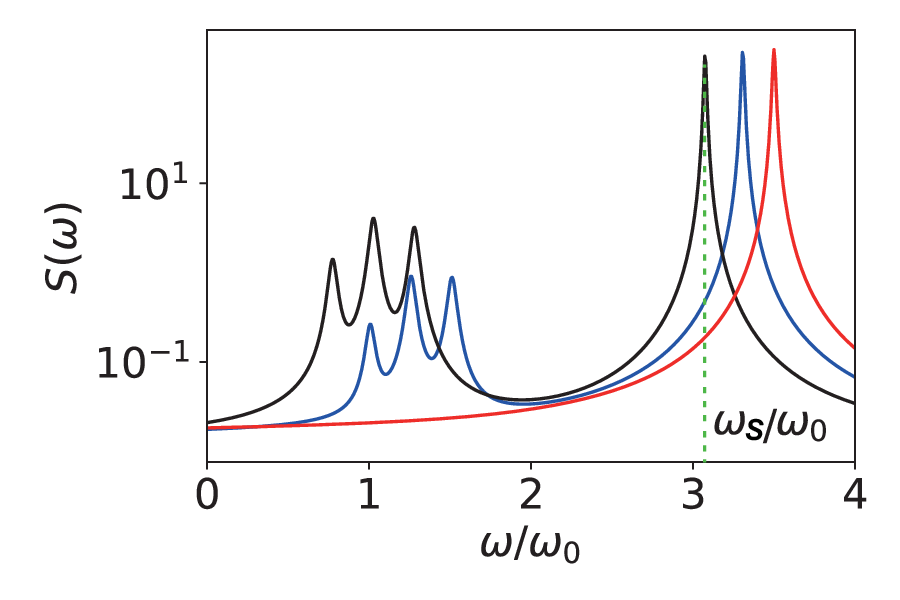}}
	\subfigure[]{\includegraphics[width=0.49\columnwidth]{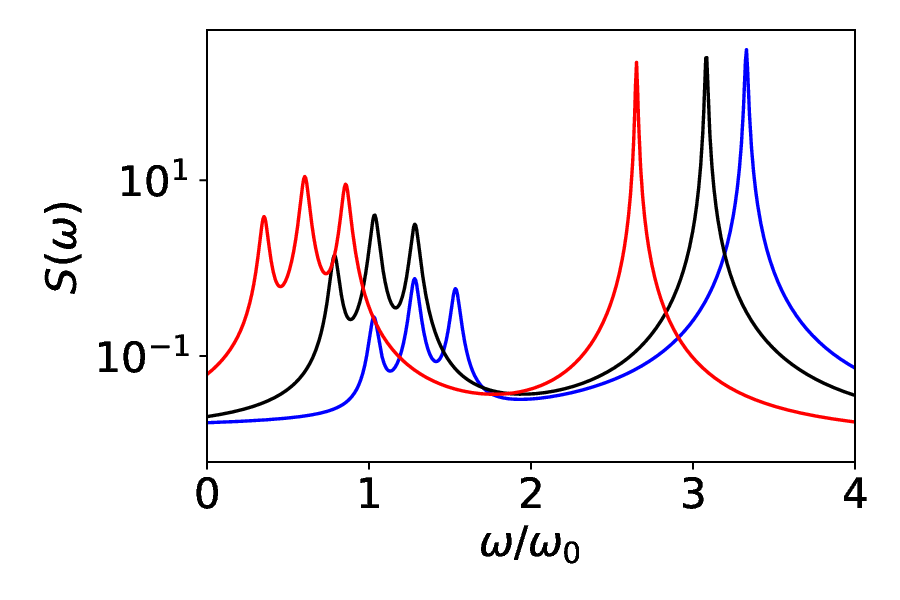}}
	\subfigure[]{\includegraphics[width=0.8\columnwidth]{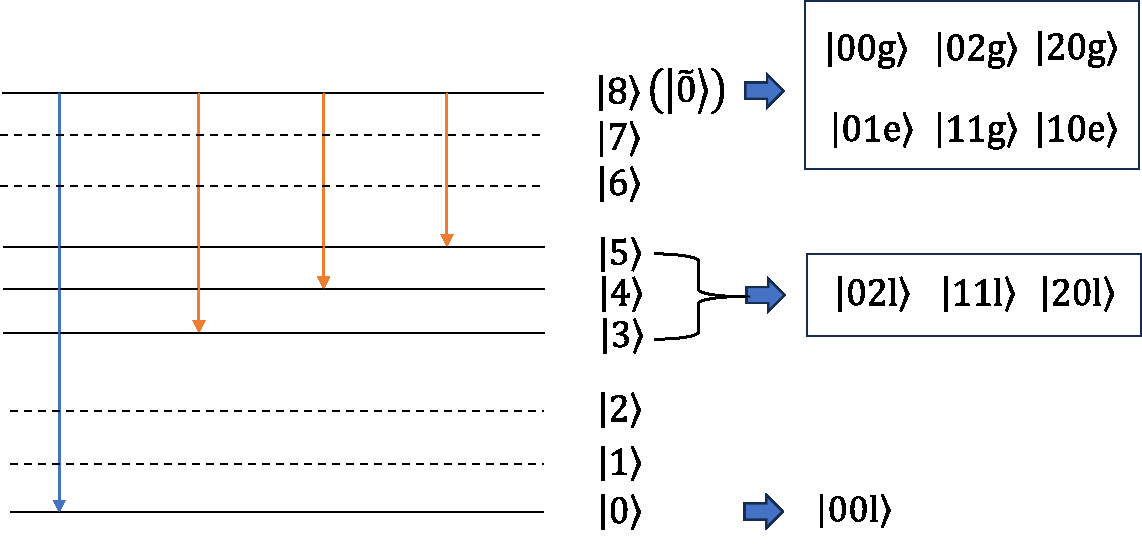}}
	\caption{Spectrum $S(\omega)$ of spontaneously emitted photons and phonons, and the energy level transition diagram. (a) The red line represents the absence of interaction between the atom and the ENZ cavity, the blue line represents the lack of coupling between the SiO$_2$ phonon mode and the atom in the cavity, and the black line represents the consideration of the interaction of the full cavity with the atom. (b) The coupling strength increases sequentially from the red line ($g_1=g_2=0.4\omega_0$) , the black line($g_1=g_2=0.6\omega_0$) to the blue line ($g_1=g_2=0.8\omega_0$). (c) Energy level transition diagram, the situation considered here is $g_1=g_2=0.6\omega_0$. In the diagram on the left Energy levels that involve transitions are shown by solid lines, and those that do not are shown by dashed lines. The bare states corresponding to the main components of each energy level of the whole system are given on the right hand side.}
	\label{specc}
\end{figure}

As mentioned previously, phonons and photons emitting from the nanocavity are caused by the spontaneous emission of the state $\left| \tilde{0} \right\rangle$. To explicitly demonstrate the mechanism,  we investigate the spectrum of photons and phonons due to the spontaneous emission of the atom
\begin{equation}
	S(\omega)=\frac{1}{2\pi}\int_{-\infty }^{\infty} \mathrm{d}t \int_{-\infty }^{\infty } \mathrm{d}t^{\prime}\left\langle\sigma^{-}(t)\sigma^{+}(t^{\prime})\right\rangle e^{-i\omega(t-t^{\prime})},
\end{equation}
where $\sigma^{+}=\sum_{j,k>j}\left\langle j\right|\sigma_{gl}+\sigma_{lg}| k \rangle \left|j\right\rangle \left\langle k \right|$ \cite{PhysRevA.88.063812, PhysRevLett.110.243601}, which is displayed in  Fig. \ref{specc}. Here we consider three different coupling cases: $g_1=g_2=0$ (red line), $g_1=0.6\omega_0, g_2=0$ (blue line), and $g_1=g_2=0.6\omega_0$ (black line). The right three highest peaks in Fig. \ref{specc} (a) at  $\omega=3.5\omega_0$ (red line), $\omega=3.307\omega_0$ (blue line) and $\omega=3.08\omega_0$ (black line) characterize the transitions from the state $\left| \tilde{0} \right\rangle$ to the ground states of the whole system in these three different cases.  Without the interaction between the atom and others (red line), the single peak spectrum at $\omega_{gl}=3.5\omega_0$  corresponds to the spontaneous emission from the atomic excited state $\left\vert g\right\rangle$ to its ground state  $\left\vert l\right\rangle$. When the interaction between the nanocavity with the atom is considered,  the three peaks will appear corresponding to the transition from $\left| \tilde{0} \right\rangle$ to $\left| 3 \right\rangle$, $\left| 4 \right\rangle$, and $\left| 5 \right\rangle$ with $\left\vert n\right\rangle$ denoting the $n$th excited state of  the whole system. The three transitions (peaks)  correspond to the frequencies $1.008\omega_0$, $1.260\omega_0$, and $1.512\omega_0$ for the blue line and to $0.782\omega_0$, $1.033\omega_0$, and $1.286\omega_0$ for the black line.
This can be further understood as follows. The parity of our system $\Pi=-\sigma_{z}\exp(i\pi N)$ with $N=a^{\dagger}a+b^{\dagger}b$ commutes with the Hamiltonian $H$, i.e., $[H,\Pi]=0$, which means the parity is conserved \cite{PhysRevA.86.033837}. The even excitations corresponding to the eigenvalue $-1$ of $\Pi$ are of odd parity, and the odd excitations corresponding to $+1$ are of even parity. Therefore, different states will have different parities. Typically, $\left| \tilde{0} \right\rangle$ is a linear combination of the bare states with even parities. Since the transitions are induced by the operators $a+a^{\dagger}$, $b+b^{\dagger}$, and $\sigma+\sigma^{\dagger}$, which will alter the parity of the state \cite{PhysRevA.94.033827}, a distinct property is that the transitions between the states in the same parity space are forbidden. Considering that the model we are discussing is a parity conserved system, the transitions can only occur between different parities. The main physical process we study is the spontaneous emission of the lowest energy state $ \left\vert \tilde{0} \right \rangle $  when there are interactions between the atom and ENZ nanocavity. Solving the eigenvalue equation by diagonalization yields that this state is mainly compsed of $\left\vert 00g \right\rangle$, which is odd parity. $\left\vert 3\right\rangle$ and $\left\vert 5\right\rangle$ are mainly composed of $\left\vert 02l \right\rangle$, $\left\vert 20l\right\rangle$, and $\left\vert 11l\right\rangle$, $\left\vert 4\right\rangle$ is composed of  $\left\vert 02l \right\rangle$ and $\left\vert 20l\right\rangle$, which are even parity. Hence, the transitions from $ \left\vert \tilde{0} \right \rangle $ to $\left\vert 3\right\rangle$, $\left\vert 4\right\rangle$, and $\left\vert 5\right\rangle$ are allowed. Meanwhile, the lowest state of the whole system $\left\vert 0 \right\rangle$ is mainly composed of $\left\vert 00l \right\rangle$, which is also even parity, and the process of photon-phonon pairs release also includes the transition from $ \left\vert \tilde{0} \right \rangle $ to $\left\vert 0 \right\rangle$. In particular, these transitions are induced by the atomic jumps and hence generate the spectrums  Fig. \ref{specc} (a).  In Fig. \ref{specc} (b), we study the influence of the coupling strength on the spectrum. The coupling strenghth does not change the parity of the system, but only affects the frequency of the transitions between different quantum states. As shown in Fig. \ref{specc} (b), the same number of peaks appear on different coupling intensity spectrum, caused by the transitions from $\left\vert \tilde{0} \right \rangle $ to $\left\vert 3\right\rangle$, $\left\vert 4\right\rangle$, $\left\vert 5\right\rangle$, and $\left\vert 0\right\rangle$. But the frequencies of the transitions corresponding to different transitions in the spectrum change with the change of the coupling strength. The schematic transition diagram and the main bare state compositions of each eigenstate of the system are represented in Fig. \ref{specc} (c).

\begin{figure}[tpb]
	\centering
	\subfigure[]{\includegraphics[width=0.49\columnwidth]{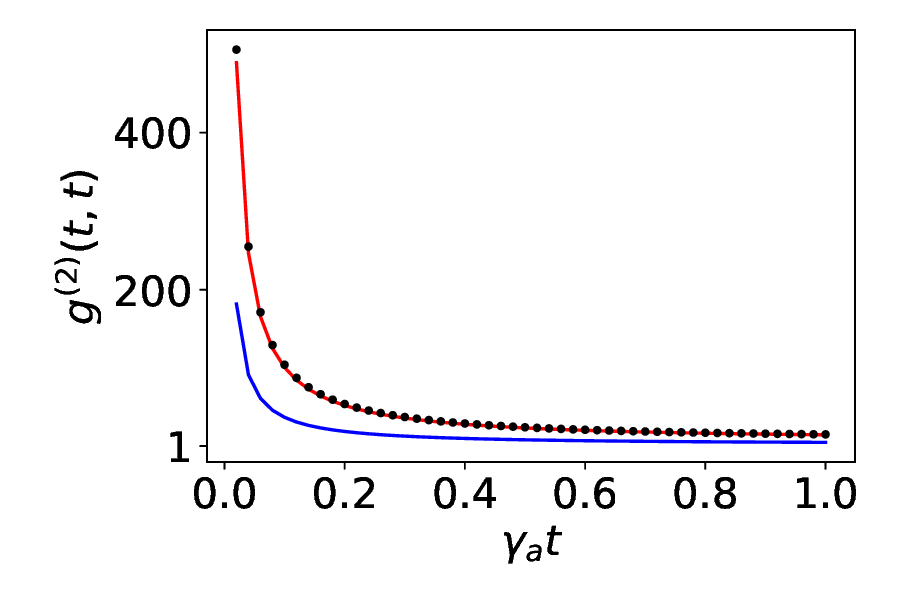}}
	\subfigure[]{\includegraphics[width=0.49\columnwidth]{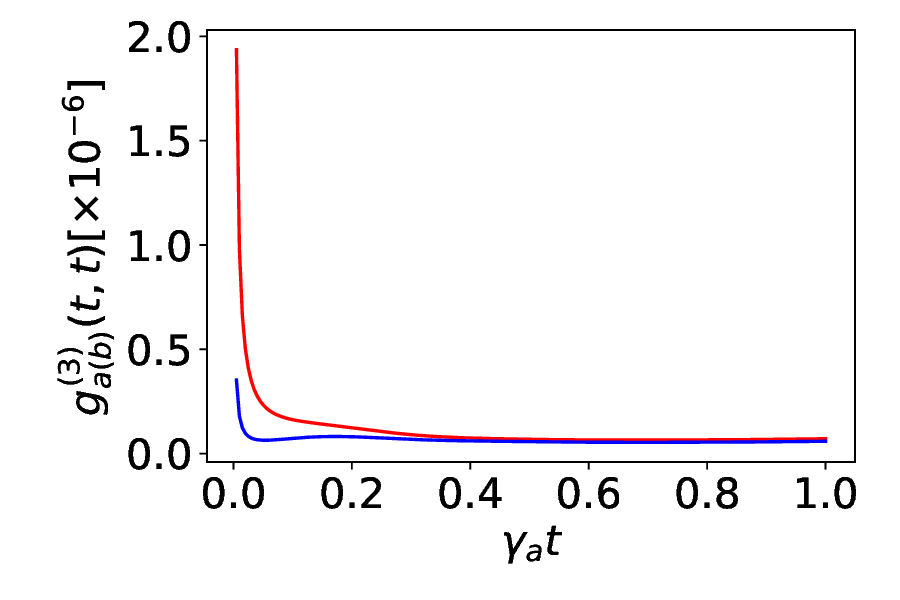}}
	\subfigure[]{\includegraphics[width=0.49\columnwidth]{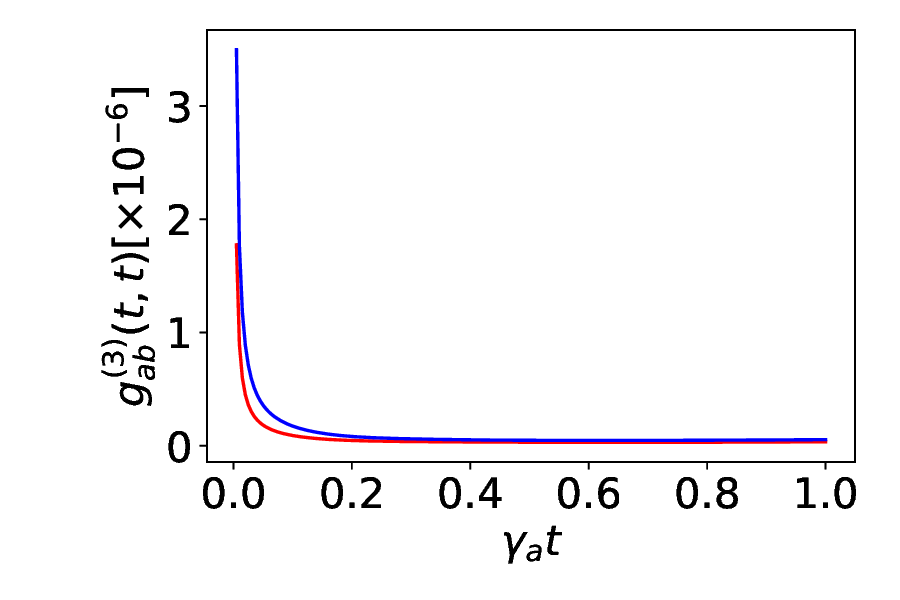}}
	\subfigure[]{\includegraphics[width=0.49\columnwidth]{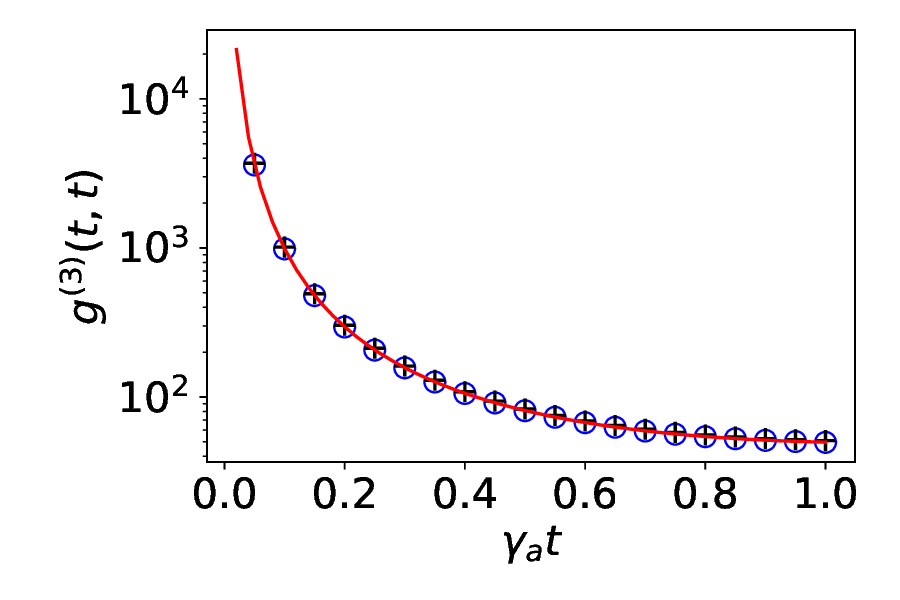}}
	\subfigure[]{\includegraphics[width=0.49\columnwidth]{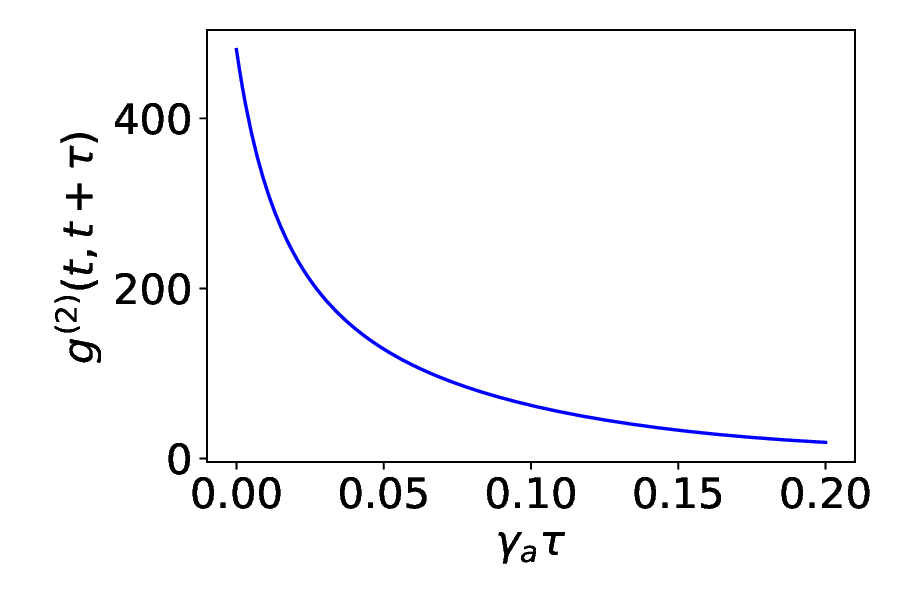}}
	\subfigure[]{\includegraphics[width=0.49\columnwidth]{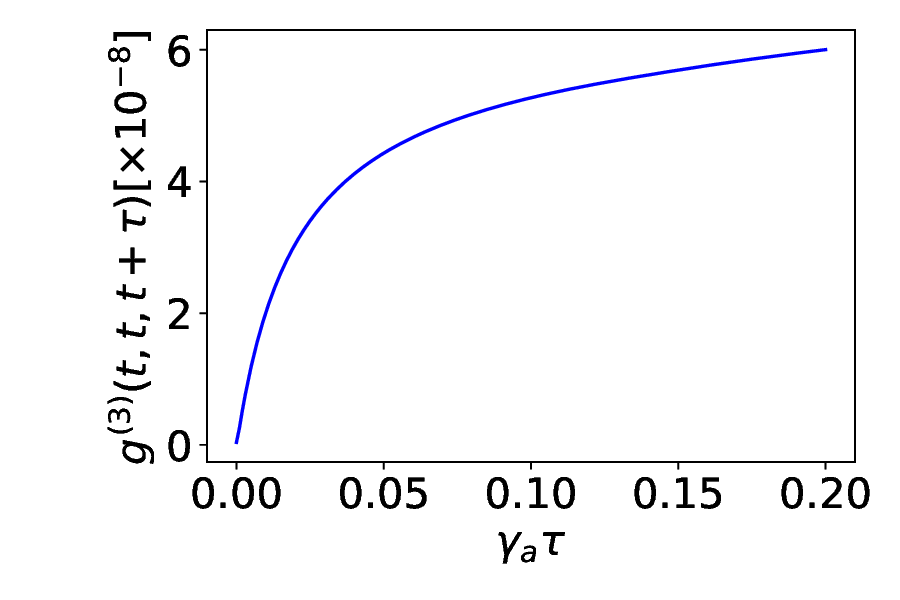}}
	\caption{Statistics of the output photons and phonons. (a) The second-order correlation function of photons and phonons $g^{(2)}(t,t)$, where the red line represents the correlation function for photons, the black dotted line for phonons, and the blue line for the second-order cross-correlation function between photons and phonons. (b) The third-order correlation function of photons and phonons $g_{a(b  )}^{(3)}(t,t)$, where the red line represents the correlation function for photons and the blue line for phonons. (c) The third-order cross-correlation function between photons and phonons $g_{ab}^{(3)}(t,t)$, the red line shows the correlation of a photon and two phonons, while the blue line shows the correlation of a phonon and two photons. (d) The third-order cross-correlation functions. The curves correspond to the probability of detecting one atomic spontaneous emitted photon accompanied by photon pairs (red), phonon pairs (blue circle), and photon-phonon pairs (black plus). (e) The second-order delay correlation function $g^{(2)}(t,t+\tau)$. (f) The third-order delay correlation function $g^{(3)}(t,t,t+\tau)$.}
	\label{corre}
\end{figure}

To better study the statistics of photon and phonon pairs emitted from the cavity, we further analyze the equal-time high-order ($n$th-order) correlation functions \cite{PhysRevLett.118.103602, PhysRevLett.109.193602}, which are defined at the moment $t$ as
\begin{equation}
	g^{(n)}(t,t)=\frac{\left\langle \prod_{k=1}^{n}X^{-}_{c_k}(t) \prod_{k=1}^{n}X^{+}_{c_k}(t)\right\rangle}{\prod_{k=1}^{n} \left\langle X^{-}_{c_k}(t) X^{+}_{c_k}(t)\right\rangle},
\end{equation}
where $c_k$ signals the operators for different modes.  For example,  $c_k=a,b$ correspond to plasmon photon mode and phonon mode, respectively, and  $c_k$ can also be designated to represent $X^{-}_{c_k}(0)=\sigma^-(0)$ for atomic operator.
The n-order correlation function can effectively characterize the statistical behavior of the particles taken into account. $g^{(n)}(t,t)>1$ indicates the n-particle bunching behavior, namely, $n$ particles have a large probability to appear together. On the contrary, there is less probability to detect $n$ particles bunching together \cite{PhysRevA.100.022501}. To show the photon/phonon pairs emitting in the system, we mainly plot the two-order Fig. \ref{corre} (a) and three-order Fig.  \ref{corre} (b) and (c) correlation functions. From the figures, one can easily find that two-order photon/phonon correlation functions and even their cross-correlation functions $g^{(2)}(t,t)>>1$, which displays the bunching behaviors of photons/phonons as well as the bunching between photons and phonons. In contrast, the various three-order correlation functions $g^{(3)}(t,t)<1$, show that one has less probability of simultaneously detecting three particles including photons and phonons.  Thus we can confirm the spontaneous release of photon/phonon pairs or photon-photon pairs. 

To further illustrate our results, we give a comparison of the equal-time and delay correlation functions \cite{PhysRevA.41.475}. The standard form of second-order delay correlation function is defined as \cite{kowalewska2019two, hamsen2017two}:
\begin{align}
	g^{(2)}(t,t+\tau)=\frac{\left\langle X^{-}_{c_k}(t) X^{-}_{c_k}(t+\tau)X^{+}_{c_k}(t+\tau) X^{+}_{c_k}(t) \right\rangle}{\left\langle X^{-}_{c_k}(t) X^{+}_{c_k}(t)\right\rangle \left\langle X^{-}_{c_k}(t+\tau) X^{+}_{c_k}(t+\tau)\right\rangle},
\end{align}
with $c_k=a,b$. Similarly, the standard form of the third-order delay correlation function is defined as:
\begin{align}
	g^{(3)}(t,t,t+\tau)=\frac{\left\langle X^{-}_{c_k}(t) X^{-}_{c_k}(t) X^{-}_{c_k}(t+\tau) X^{+}_{c_k}(t+\tau)X^{+}_{c_k}(t) X^{+}_{c_k}(t) \right\rangle}{\left\langle X^{-}_{c_k}(t) X^{+}_{c_k}(t)\right\rangle^2 \left\langle X^{-}_{c_k}(t+\tau) X^{+}_{c_k}(t+\tau)\right\rangle},
\end{align}

As shown in Fig. \ref{corre}(e)-(f), we give a comparison between the delay correlation function and the equal-time correlation function. We found that $g^{(2)}(t,t)>g^{(2)}(t,t+\tau)$, $g^{(3)}(t,t,t)<g^{(3)}(t,t,t+\tau)$, which is further shown that photon-phonon pairs are antibunching with each other and photons and phonons are produced in pairs. Here we take the initial time is $t=\gamma_a$.
 Meanwhile, we also calculated the three-order cross-correlation functions of atomic spontaneous emission and photon/phonon pairs in Fig. \ref{corre} (d). The three correlation functions are apparently much larger than 1, which further means the large probability of detecting one atomic spontaneous emitted photon (corresponding to the three lower peaks in Fig. \ref{specc} (a) accompanied by photon pairs, phonon pairs, and photon-phonon pairs. 

Finally, we'd like to emphasize that the ultrastrong coupling of the plasmon mode and the phonon has been experimentally achieved \cite{Nat.Photonics15}. With the development of the experiment, the coupling strength of phonons, photons, and atoms keeps increasing \cite{s41567-019-0534-4, J.Phys.Chem.Lett.11.7, PhysRevLett.123.247701, s41467-020-16524-x}, entering the USC regime. Meanwhile, the experimental detection of phonon pairs and photon pairs can be performed by homodyne detection \cite{PhysRevA.55.3117}, optical interference detection \cite{Li:22}, Raman spectroscopy \cite{Nanoscale2022.14}, etc. So our scheme should be feasible for generating phonon pairs and photon pairs in this new ENZ nanocavity \cite{Nat.Photonics15}. The parameters selected in our calculations are experimentally feasible: vibration constant $\omega_p=0.25\omega_0$ \cite{Nat.Photonics15}, $g\approx \omega_0, \omega_b(0.2<g/\omega_0,\omega_b<1)$ in the USC regime \cite{nphys1730}, and dissipation parameters $\gamma_a, \gamma_b, \gamma_{eg}, \gamma_{gl}/\omega_0 \approx 10^{-2}$ are shown in the Methods section on page 7 of this work \cite{Nat.Commun.12.6206}. In our work we use the frequency $\omega_0/2\pi=2.782$GHZ used in Figure 1 of this paper \cite{nphys1730}, and the adjustment of the coupling constant and dissipation frequency in our study are in the range of the above experimental papers. We have verified that the physical properties obtained are independent of the specific values of the parameters when the parameter ratios are the same. In addition, all the results in this article are calculated numerically using Qutip \cite{Comp.Phys.Comm.184(2013)1234}, where we truncate the numbers of photons and phonons up to 4. We find that the lowest state of energy with interactions and the energy levels below are almost independent of the truncation dimension, so the physical processes we study are not subject to change with the truncation dimension. We have verified that the final results of this study are almost not affected by the higher dimension of Hilbert space.

\section{Conclusion}
We have investigated the detection of the photon and phonon pairs in the ground state of the qubit-plasmon-phonon hybrid system in the USC regime. The system includes an artificial three-level atom with only the upper two energy levels coupled to plasmon mode and SiO$_2$ phonon, and the transition frequency of the first and the second excited state is resonant with the photon and phonon. In this way, the upper and lower energy levels can be regarded as a two-level system. It can be found that the spontaneous emission of the atom from the intermediate state to the ground state is accompanied by the generation of photon pairs and phonon pairs. The detected photon and phonon number rates have increased compared to before, which is easier to detect in experiments. It is worth noting that the coupling of the phonon and the atom in the ENZ nanocavity we studied can promote the maximum value of the output photon and phonon number rates. In addition, we also analyzed the spectrum $S(\omega)$ of the emitted photons and phonons and compared the spectrum with and without interaction. Finally, by calculating the correlation function of photons and phonons, we confirmed that photons and phonons are emitted in pairs. However, the experimental realization of the ultrastrong coupling between the three-level atom and the ENZ cavity is also briefly discussed. Our research will advance the study of the ground state properties in the USC regime.

\section*{Acknowledgements}
This work was supported by the National Natural Science
Foundation of China under Grant No.12175029, No.11775040, and No. 12011530014.

\section*{Disclosures}
The authors declare no conflicts of interest.

\section*{Data availability}
No data were generated or analyzed in the presented research.

\bibliography{t}

\begin{thebibliography}{10}
\newcommand{\enquote}[1]{``#1''}

\bibitem{10.1021/acs.nanolett.7b03103}
A.~Bayer, M.~Pozimski, S.~Schambeck, D.~Schuh, R.~Huber, D.~Bougeard, and
  C.~Lange, \enquote{Terahertz light–matter interaction beyond unity coupling
  strength,} {\protect\JournalTitle{Nano Letters}} \textbf{17}, 6340--6344
  (2017).

\bibitem{doi:10.1038/nphys3906}
F.~Yoshihara, T.~Fuse, S.~Ashhab, K.~Kakuyanagi, S.~Saito, and K.~Semba,
  \enquote{Superconducting qubit--oscillator circuit beyond the
  ultrastrong-coupling regime,} {\protect\JournalTitle{Nature Physics}}
  \textbf{13}, 44--47 (2017).

\bibitem{PhysRevLett.105.237001}
P.~Forn-D\'{\i}az, J.~Lisenfeld, D.~Marcos, J.~J. Garc\'{\i}a-Ripoll,
  E.~Solano, C.~J. P.~M. Harmans, and J.~E. Mooij, \enquote{Observation of the
  bloch-siegert shift in a qubit-oscillator system in the ultrastrong coupling
  regime,} {\protect\JournalTitle{Phys. Rev. Lett.}} \textbf{105}, 237001
  (2010).

\bibitem{PhysRevB.93.214501}
A.~Baust, E.~Hoffmann, M.~Haeberlein, M.~J. Schwarz, P.~Eder, J.~Goetz,
  F.~Wulschner, E.~Xie, L.~Zhong, F.~Quijandr\'{\i}a, D.~Zueco, J.-J.~G.
  Ripoll, L.~Garc\'{\i}a-\'Alvarez, G.~Romero, E.~Solano, K.~G. Fedorov, E.~P.
  Menzel, F.~Deppe, A.~Marx, and R.~Gross, \enquote{{Ultrastrong coupling in
  two-resonator circuit QED},} {\protect\JournalTitle{Phys. Rev. B}}
  \textbf{93}, 214501 (2016).

\bibitem{Nat.Photonics15}
D.~Yoo, F.~de~Le{\'o}n-P{\'e}rez, M.~Pelton, I.-H. Lee, D.~A. Mohr, M.~B.
  Raschke, J.~D. Caldwell, L.~Mart{\'\i}n-Moreno, and S.-H. Oh,
  \enquote{Ultrastrong plasmon--phonon coupling via epsilon-near-zero
  nanocavities,} {\protect\JournalTitle{Nature Photonics}} \textbf{15},
  125--130 (2021).

\bibitem{10.1038/nphys3905}
P.~Forn-D{\'\i}az, J.~J. Garc{\'\i}a-Ripoll, B.~Peropadre, J.-L. Orgiazzi,
  M.~Yurtalan, R.~Belyansky, C.~M. Wilson, and A.~Lupascu, \enquote{Ultrastrong
  coupling of a single artificial atom to an electromagnetic continuum in the
  nonperturbative regime,} {\protect\JournalTitle{Nature Physics}} \textbf{13},
  39--43 (2017).

\bibitem{PhysRevLett.120.093602}
C.~Leroux, L.~C.~G. Govia, and A.~A. Clerk, \enquote{Enhancing cavity quantum
  electrodynamics via antisqueezing: Synthetic ultrastrong coupling,}
  {\protect\JournalTitle{Phys. Rev. Lett.}} \textbf{120}, 093602 (2018).

\bibitem{PhysRevLett.109.193602}
A.~Ridolfo, M.~Leib, S.~Savasta, and M.~J. Hartmann, \enquote{Photon blockade
  in the ultrastrong coupling regime,} {\protect\JournalTitle{Phys. Rev.
  Lett.}} \textbf{109}, 193602 (2012).

\bibitem{PhysRevResearch.4.013013}
K.~Koshino, T.~Shitara, Z.~Ao, and K.~Semba, \enquote{{Deterministic
  three-photon down-conversion by a passive ultrastrong cavity-QED system},}
  {\protect\JournalTitle{Phys. Rev. Research}} \textbf{4}, 013013 (2022).

\bibitem{npj.QuantumInf.3.46(2017)}
S.~J. Bosman, M.~F. Gely, V.~Singh, A.~Bruno, D.~Bothner, and G.~A. Steele,
  \enquote{Multi-mode ultra-strong coupling in circuit quantum
  electrodynamics,} {\protect\JournalTitle{npj Quantum Information}}
  \textbf{3}, 46 (2017).

\bibitem{PhysRevLett.127.073602}
Q.~Bin, Y.~Wu, and X.-Y. L\"u, \enquote{Parity-symmetry-protected multiphoton
  bundle emission,} {\protect\JournalTitle{Phys. Rev. Lett.}} \textbf{127},
  073602 (2021).

\bibitem{PhysRevA.105.023720}
V.~Macr\'{\i}, F.~Minganti, A.~F. Kockum, A.~Ridolfo, S.~Savasta, and F.~Nori,
  \enquote{Revealing higher-order light and matter energy exchanges using
  quantum trajectories in ultrastrong coupling,} {\protect\JournalTitle{Phys.
  Rev. A}} \textbf{105}, 023720 (2022).

\bibitem{New.J.Phys.18.123005}
R.~Stassi, S.~Savasta, L.~Garziano, B.~Spagnolo, and F.~Nori, \enquote{{Output
  field-quadrature measurements and squeezing in ultrastrong cavity-QED},}
  {\protect\JournalTitle{New J. Phys}} \textbf{18}, 123005 (2016).

\bibitem{PhysRevResearch.4.023016}
M.~Wang, S.~Mallick, A.~F. Kockum, and K.~B\"orjesson, \enquote{Organic charged
  polaritons in the ultrastrong coupling regime,} {\protect\JournalTitle{Phys.
  Rev. Research}} \textbf{4}, 023016 (2022).

\bibitem{PhysRevA.98.053834}
A.~Settineri, V.~Macr\'{\i}, A.~Ridolfo, O.~Di~Stefano, A.~F. Kockum, F.~Nori,
  and S.~Savasta, \enquote{Dissipation and thermal noise in hybrid quantum
  systems in the ultrastrong-coupling regime,} {\protect\JournalTitle{Phys.
  Rev. A}} \textbf{98}, 053834 (2018).

\bibitem{PhysRevB.101.214414}
J.~R. Everts, G.~G.~G. King, N.~J. Lambert, S.~Kocsis, S.~Rogge, and J.~J.
  Longdell, \enquote{Ultrastrong coupling between a microwave resonator and
  antiferromagnetic resonances of rare-earth ion spins,}
  {\protect\JournalTitle{Phys. Rev. B}} \textbf{101}, 214414 (2020).

\bibitem{PhysRevA.81.042311}
S.~Ashhab and F.~Nori, \enquote{Qubit-oscillator systems in the
  ultrastrong-coupling regime and their potential for preparing nonclassical
  states,} {\protect\JournalTitle{Phys. Rev. A}} \textbf{81}, 042311 (2010).

\bibitem{PhysRevA.95.063849}
A.~F. Kockum, A.~Miranowicz, V.~Macr\`{\i}, S.~Savasta, and F.~Nori,
  \enquote{Deterministic quantum nonlinear optics with single atoms and virtual
  photons,} {\protect\JournalTitle{Phys. Rev. A}} \textbf{95}, 063849 (2017).

\bibitem{PhysRevLett.110.243601}
R.~Stassi, A.~Ridolfo, O.~Di~Stefano, M.~J. Hartmann, and S.~Savasta,
  \enquote{Spontaneous conversion from virtual to real photons in the
  ultrastrong-coupling regime,} {\protect\JournalTitle{Phys. Rev. Lett.}}
  \textbf{110}, 243601 (2013).

\bibitem{RevModPhys.91.025005}
P.~Forn-D\'{\i}az, L.~Lamata, E.~Rico, J.~Kono, and E.~Solano,
  \enquote{Ultrastrong coupling regimes of light-matter interaction,}
  {\protect\JournalTitle{Rev. Mod. Phys.}} \textbf{91}, 025005 (2019).

\bibitem{Nat.Commun.8.1715(2017)}
N.~K. Langford, R.~Sagastizabal, M.~Kounalakis, C.~Dickel, A.~Bruno, F.~Luthi,
  D.~J. Thoen, A.~Endo, and L.~DiCarlo, \enquote{Experimentally simulating the
  dynamics of quantum light and matter at deep-strong coupling,}
  {\protect\JournalTitle{Nature communications}} \textbf{8}, 1715 (2017).

\bibitem{PhysRevLett.116.113601}
M.~Cirio, S.~De~Liberato, N.~Lambert, and F.~Nori, \enquote{Ground state
  electroluminescence,} {\protect\JournalTitle{Phys. Rev. Lett.}} \textbf{116},
  113601 (2016).

\bibitem{Adv.Quantum.Technol.3.12}
M.-S. Choi, \enquote{Exotic quantum states of circuit quantum electrodynamics
  in the ultra-strong coupling regime,} {\protect\JournalTitle{Adv. Quantum
  Technol.}} \textbf{3}, 2000085 (2020).

\bibitem{PhysRevA.84.043832}
F.~Beaudoin, J.~M. Gambetta, and A.~Blais, \enquote{{Dissipation and
  ultrastrong coupling in circuit QED},} {\protect\JournalTitle{Phys. Rev. A}}
  \textbf{84}, 043832 (2011).

\bibitem{nphys1730}
T.~Niemczyk, F.~Deppe, H.~Huebl, E.~Menzel, F.~Hocke, M.~Schwarz,
  J.~Garcia-Ripoll, D.~Zueco, T.~H{\"u}mmer, E.~Solano, A.~Marx, and R.~Gross,
  \enquote{Circuit quantum electrodynamics in the ultrastrong-coupling regime,}
  {\protect\JournalTitle{Nature Physics}} \textbf{6}, 772--776 (2010).

\bibitem{PhysRevA.105.053718}
T.-t. Ma, D.~B. Horoshko, C.-s. Yu, and S.~Y. Kilin, \enquote{Photon and phonon
  statistics in a qubit-plasmon-phonon ultrastrong-coupling system,}
  {\protect\JournalTitle{Phys. Rev. A}} \textbf{105}, 053718 (2022).

\bibitem{PhysRevD.100.125019}
J.-T. Hsiang and B.-L. Hu, \enquote{Ground state excitation of an atom strongly
  coupled to a free quantum field,} {\protect\JournalTitle{Phys. Rev. D}}
  \textbf{100}, 125019 (2019).

\bibitem{PhysRevLett.114.183601}
J.~Lolli, A.~Baksic, D.~Nagy, V.~E. Manucharyan, and C.~Ciuti,
  \enquote{Ancillary qubit spectroscopy of vacua in cavity and circuit quantum
  electrodynamics,} {\protect\JournalTitle{Phys. Rev. Lett.}} \textbf{114},
  183601 (2015).

\bibitem{PhysRevA.100.062501}
W.~Qin, V.~Macr\`{\i}, A.~Miranowicz, S.~Savasta, and F.~Nori,
  \enquote{Emission of photon pairs by mechanical stimulation of the squeezed
  vacuum,} {\protect\JournalTitle{Phys. Rev. A}} \textbf{100}, 062501 (2019).

\bibitem{PhysRevA.96.012325}
Z.~Chen, Y.~Wang, T.~Li, L.~Tian, Y.~Qiu, K.~Inomata, F.~Yoshihara, S.~Han,
  F.~Nori, J.~S. Tsai, and J.~Q. You, \enquote{Single-photon-driven high-order
  sideband transitions in an ultrastrongly coupled
  circuit-quantum-electrodynamics system,} {\protect\JournalTitle{Phys. Rev.
  A}} \textbf{96}, 012325 (2017).

\bibitem{Journal.of.Applied.Physics.113.136510(2013)}
G.~Scalari, C.~Maissen, D.~Hagenm{\"u}ller, S.~De~Liberato, C.~Ciuti,
  C.~Reichl, W.~Wegscheider, D.~Schuh, M.~Beck, and J.~Faist,
  \enquote{Ultrastrong light-matter coupling at terahertz frequencies with
  split ring resonators and inter-landau level transitions,}
  {\protect\JournalTitle{Journal of Applied Physics}} \textbf{113} (2013).

\bibitem{PhysRevB.90.205309}
C.~Maissen, G.~Scalari, F.~Valmorra, M.~Beck, J.~Faist, S.~Cibella, R.~Leoni,
  C.~Reichl, C.~Charpentier, and W.~Wegscheider, \enquote{Ultrastrong coupling
  in the near field of complementary split-ring resonators,}
  {\protect\JournalTitle{Phys. Rev. B}} \textbf{90}, 205309 (2014).

\bibitem{10.1038/nphys3850}
Q.~Zhang, M.~Lou, X.~Li, J.~L. Reno, W.~Pan, J.~D. Watson, M.~J. Manfra, and
  J.~Kono, \enquote{Collective non-perturbative coupling of 2d electrons with
  high-quality-factor terahertz cavity photons,} {\protect\JournalTitle{Nature
  Physics}} \textbf{12}, 1005--1011 (2016).

\bibitem{PhysRevB.83.075309}
V.~M. Muravev, I.~V. Andreev, I.~V. Kukushkin, S.~Schmult, and W.~Dietsche,
  \enquote{Observation of hybrid plasmon-photon modes in microwave transmission
  of coplanar microresonators,} {\protect\JournalTitle{Phys. Rev. B}}
  \textbf{83}, 075309 (2011).

\bibitem{PhysRevB.101.075301}
J.~Keller, G.~Scalari, F.~Appugliese, S.~Rajabali, M.~Beck, J.~Haase, C.~A.
  Lehner, W.~Wegscheider, M.~Failla, M.~Myronov, D.~R. Leadley,
  J.~Lloyd-Hughes, P.~Nataf, and J.~Faist, \enquote{Landau polaritons in highly
  nonparabolic two-dimensional gases in the ultrastrong coupling regime,}
  {\protect\JournalTitle{Phys. Rev. B}} \textbf{101}, 075301 (2020).

\bibitem{PhysRevA.96.022308}
S.-A. Biehs and G.~S. Agarwal, \enquote{Qubit entanglement across
  $\ensuremath{\epsilon}$-near-zero media,} {\protect\JournalTitle{Phys. Rev.
  A}} \textbf{96}, 022308 (2017).

\bibitem{acs.nanolett.8b04182}
E.~L. Runnerstrom, K.~P. Kelley, T.~G. Folland, J.~R. Nolen, N.~Engheta, J.~D.
  Caldwell, and J.-P. Maria, \enquote{Polaritonic hybrid-epsilon-near-zero
  modes: beating the plasmonic confinement vs propagation-length trade-off with
  doped cadmium oxide bilayers,} {\protect\JournalTitle{Nano letters}}
  \textbf{19}, 948--957 (2018).

\bibitem{Rodrigo:21}
S.~G. Rodrigo, \enquote{Amplification of stimulated light emission in arrays of
  nanoholes by plasmonic absorption-induced transparency,}
  {\protect\JournalTitle{Optics Express}} \textbf{29}, 30715--30726 (2021).

\bibitem{nanoph-2020-0449}
D.~N. Basov, A.~Asenjo-Garcia, P.~J. Schuck, X.~Zhu, and A.~Rubio,
  \enquote{Polariton panorama,} {\protect\JournalTitle{Nanophotonics}}
  \textbf{10}, 549--577 (2021).

\bibitem{science.aau7742}
A.~Thomas, L.~Lethuillier-Karl, K.~Nagarajan, R.~M. Vergauwe, J.~George,
  T.~Chervy, A.~Shalabney, E.~Devaux, C.~Genet, J.~Moran, and E.~T. W.,
  \enquote{Tilting a ground-state reactivity landscape by vibrational strong
  coupling,} {\protect\JournalTitle{Science}} \textbf{363}, 615--619 (2019).

\bibitem{PhysRevLett.110.163601}
A.~Ridolfo, S.~Savasta, and M.~J. Hartmann, \enquote{Nonclassical radiation
  from thermal cavities in the ultrastrong coupling regime,}
  {\protect\JournalTitle{Phys. Rev. Lett.}} \textbf{110}, 163601 (2013).

\bibitem{PhysRevLett.117.043601}
L.~Garziano, V.~Macr\`{\i}, R.~Stassi, O.~Di~Stefano, F.~Nori, and S.~Savasta,
  \enquote{One photon can simultaneously excite two or more atoms,}
  {\protect\JournalTitle{Phys. Rev. Lett.}} \textbf{117}, 043601 (2016).

\bibitem{PhysRevA.99.033809}
Q.~Bin, X.-Y. L\"u, T.-S. Yin, Y.~Li, and Y.~Wu, \enquote{Collective radiance
  effects in the ultrastrong-coupling regime,} {\protect\JournalTitle{Phys.
  Rev. A}} \textbf{99}, 033809 (2019).

\bibitem{Adv.Quantum.Technol.3.7}
A.~L. Boit\'{e}, \enquote{Theoretical methods for ultrastrong light-matter
  interactions,} {\protect\JournalTitle{Adv. Quantum Technol.}} \textbf{3},
  1900140 (2020).

\bibitem{PhysRevLett.126.153603}
Y.~Ashida, A.~m.~c. \ifmmode \dot{I}\else \.{I}\fi{}mamo\ifmmode~\breve{g}\else
  \u{g}\fi{}lu, and E.~Demler, \enquote{Cavity quantum electrodynamics at
  arbitrary light-matter coupling strengths,} {\protect\JournalTitle{Phys. Rev.
  Lett.}} \textbf{126}, 153603 (2021).

\bibitem{booksee.org/book/1397869}
H.~P. Breuer and F.~Petruccione, \emph{The Theory of Open Quantum Systems}
  (Oxford University, New York, 2006).

\bibitem{PhysRevLett.120.183601}
F.~Yoshihara, T.~Fuse, Z.~Ao, S.~Ashhab, K.~Kakuyanagi, S.~Saito, T.~Aoki,
  K.~Koshino, and K.~Semba, \enquote{Inversion of qubit energy levels in
  qubit-oscillator circuits in the deep-strong-coupling regime,}
  {\protect\JournalTitle{Phys. Rev. Lett.}} \textbf{120}, 183601 (2018).

\bibitem{PhysRevA.31.3761}
C.~W. Gardiner and M.~J. Collett, \enquote{Input and output in damped quantum
  systems: Quantum stochastic differential equations and the master equation,}
  {\protect\JournalTitle{Phys. Rev. A}} \textbf{31}, 3761--3774 (1985).

\bibitem{PhysRevA.74.033811}
C.~Ciuti and I.~Carusotto, \enquote{Input-output theory of cavities in the
  ultrastrong coupling regime: The case of time-independent cavity parameters,}
  {\protect\JournalTitle{Phys. Rev. A}} \textbf{74}, 033811 (2006).

\bibitem{PhysRevA.102.043701}
J.~Li, C.~Ding, and Y.~Wu, \enquote{Strongly correlated photons with quantum
  feedback in a cascaded nanoscale double-cavity system,}
  {\protect\JournalTitle{Phys. Rev. A}} \textbf{102}, 043701 (2020).

\bibitem{s41598-018-36056-1}
O.~Di~Stefano, A.~F. Kockum, A.~Ridolfo, S.~Savasta, and F.~Nori,
  \enquote{Photodetection probability in quantum systems with arbitrarily
  strong light-matter interaction,} {\protect\JournalTitle{Scientific Reports}}
  \textbf{8}, 17825 (2018).

\bibitem{s41467-017-01504-5}
S.~D. Liberato, \enquote{Virtual photons in the ground state of a dissipative
  system,} {\protect\JournalTitle{Nat Commun}} \textbf{8}, 1465 (2017).

\bibitem{PhysRevA.89.033827}
J.-F. Huang and C.~K. Law, \enquote{Photon emission via vacuum-dressed
  intermediate states under ultrastrong coupling,} {\protect\JournalTitle{Phys.
  Rev. A}} \textbf{89}, 033827 (2014).

\bibitem{10.1038/nature11821}
J.-M. Pirkkalainen, S.~Cho, J.~Li, G.~Paraoanu, P.~Hakonen, and
  M.~Sillanp{\"a}{\"a}, \enquote{Hybrid circuit cavity quantum electrodynamics
  with a micromechanical resonator,} {\protect\JournalTitle{Nature}}
  \textbf{494}, 211--215 (2013).

\bibitem{s41586-019-0960-6}
C.~Fl{\"u}hmann, T.~L. Nguyen, M.~Marinelli, V.~Negnevitsky, K.~Mehta, and
  J.~Home, \enquote{Encoding a qubit in a trapped-ion mechanical oscillator,}
  {\protect\JournalTitle{Nature}} \textbf{566}, 513--517 (2019).

\bibitem{PhysRevB.72.115303}
C.~Ciuti, G.~Bastard, and I.~Carusotto, \enquote{Quantum vacuum properties of
  the intersubband cavity polariton field,} {\protect\JournalTitle{Phys. Rev.
  B}} \textbf{72}, 115303 (2005).

\bibitem{PhysRevA.104.023109}
Y.~Wang and S.~D. Liberato, \enquote{Theoretical proposals to measure
  resonator-induced modifications of the electronic ground state in doped
  quantum wells,} {\protect\JournalTitle{Phys. Rev. A}} \textbf{104}, 023109
  (2021).

\bibitem{PhysRevA.100.063827}
X.~Wang, W.~Qin, A.~Miranowicz, S.~Savasta, and F.~Nori,
  \enquote{Unconventional cavity optomechanics: Nonlinear control of phonons in
  the acoustic quantum vacuum,} {\protect\JournalTitle{Phys. Rev. A}}
  \textbf{100}, 063827 (2019).

\bibitem{NOH2019350}
C.~Noh and H.~Nha, \enquote{Output field squeezing in a weakly-driven
  dissipative quantum rabi model,} {\protect\JournalTitle{Optics
  Communications}} \textbf{435}, 350--354 (2019).

\bibitem{AIP.Advances.10.025106}
D.~Manzano, \enquote{A short introduction to the lindblad master equation,}
  {\protect\JournalTitle{AIP Advances}} \textbf{10}, 025106 (2020).

\bibitem{ref2004Gemma}
H.-P. Breuer and F.~Petruccione, \emph{The theory of open quantum systems}
  (Oxford University Press, Oxford, 2002).

\bibitem{Horoshko:98}
D.~B. Horoshko and S.~Y. Kilin, \enquote{Multimode unraveling of master
  equation and decoherence problem,} {\protect\JournalTitle{Opt. Express}}
  \textbf{2}, 347--354 (1998).

\bibitem{PhysRevLett.105.133601}
M.~Mariantoni, E.~P. Menzel, F.~Deppe, M.~A. Araque~Caballero, A.~Baust,
  T.~Niemczyk, E.~Hoffmann, E.~Solano, A.~Marx, and R.~Gross, \enquote{Planck
  spectroscopy and quantum noise of microwave beam splitters,}
  {\protect\JournalTitle{Phys. Rev. Lett.}} \textbf{105}, 133601 (2010).

\bibitem{10.1038/nphys1845}
D.~Bozyigit, C.~Lang, L.~Steffen, J.~Fink, C.~Eichler, M.~Baur, R.~Bianchetti,
  P.~J. Leek, S.~Filipp, M.~P. Da~Silva, A.~Blais, and A.~Wallraff,
  \enquote{Antibunching of microwave-frequency photons observed in correlation
  measurements using linear detectors,} {\protect\JournalTitle{Nature Physics}}
  \textbf{7}, 154--158 (2011).

\bibitem{PhysRevLett.105.100401}
E.~P. Menzel, F.~Deppe, M.~Mariantoni, M.~A. Araque~Caballero, A.~Baust,
  T.~Niemczyk, E.~Hoffmann, A.~Marx, E.~Solano, and R.~Gross,
  \enquote{Dual-path state reconstruction scheme for propagating quantum
  microwaves and detector noise tomography,} {\protect\JournalTitle{Phys. Rev.
  Lett.}} \textbf{105}, 100401 (2010).

\bibitem{PhysRevA.98.053811}
F.~Graffitti, J.~Kelly-Massicotte, A.~Fedrizzi, and A.~M.
  Bra\ifmmode~\acute{n}\else \'{n}\fi{}czyk, \enquote{Design considerations for
  high-purity heralded single-photon sources,} {\protect\JournalTitle{Phys.
  Rev. A}} \textbf{98}, 053811 (2018).

\bibitem{PhysRevA.88.063812}
A.~Ridolfo, E.~del Valle, and M.~J. Hartmann, \enquote{Photon correlations from
  ultrastrong optical nonlinearities,} {\protect\JournalTitle{Phys. Rev. A}}
  \textbf{88}, 063812 (2013).

\bibitem{PhysRevA.86.033837}
S.~He, C.~Wang, Q.-H. Chen, X.-Z. Ren, T.~Liu, and K.-L. Wang,
  \enquote{First-order corrections to the rotating-wave approximation in the
  jaynes-cummings model,} {\protect\JournalTitle{Phys. Rev. A}} \textbf{86},
  033837 (2012).

\bibitem{PhysRevA.94.033827}
A.~Le~Boit\'e, M.-J. Hwang, H.~Nha, and M.~B. Plenio, \enquote{Fate of photon
  blockade in the deep strong-coupling regime,} {\protect\JournalTitle{Phys.
  Rev. A}} \textbf{94}, 033827 (2016).

\bibitem{PhysRevLett.118.103602}
J.~Goetz, S.~Pogorzalek, F.~Deppe, K.~G. Fedorov, P.~Eder, M.~Fischer,
  F.~Wulschner, E.~Xie, A.~Marx, and R.~Gross, \enquote{Photon statistics of
  propagating thermal microwaves,} {\protect\JournalTitle{Phys. Rev. Lett.}}
  \textbf{118}, 103602 (2017).

\bibitem{PhysRevA.100.022501}
A.~Settineri, V.~Macr\`{\i}, L.~Garziano, O.~Di~Stefano, F.~Nori, and
  S.~Savasta, \enquote{Conversion of mechanical noise into correlated photon
  pairs: Dynamical casimir effect from an incoherent mechanical drive,}
  {\protect\JournalTitle{Phys. Rev. A}} \textbf{100}, 022501 (2019).

\bibitem{PhysRevA.41.475}
X.~T. Zou and L.~Mandel, \enquote{Photon-antibunching and sub-poissonian photon
  statistics,} {\protect\JournalTitle{Phys. Rev. A}} \textbf{41}, 475--476
  (1990).

\bibitem{kowalewska2019two}
A.~Kowalewska-Kud{\l}aszyk, S.~I. Abo, G.~Chimczak, J.~Pe{\v{r}}ina~Jr,
  F.~Nori, and A.~Miranowicz, \enquote{Two-photon blockade and photon-induced
  tunneling generated by squeezing,} {\protect\JournalTitle{Physical Review A}}
  \textbf{100}, 053857 (2019).

\bibitem{hamsen2017two}
C.~Hamsen, K.~N. Tolazzi, T.~Wilk, and G.~Rempe, \enquote{Two-photon blockade
  in an atom-driven cavity qed system,} {\protect\JournalTitle{Physical review
  letters}} \textbf{118}, 133604 (2017).

\bibitem{s41567-019-0534-4}
O.~Di~Stefano, A.~Settineri, V.~Macr{\`\i}, L.~Garziano, R.~Stassi, S.~Savasta,
  and F.~Nori, \enquote{Resolution of gauge ambiguities in ultrastrong-coupling
  cavity quantum electrodynamics,} {\protect\JournalTitle{Nature Physics}}
  \textbf{15}, 803--808 (2019).

\bibitem{J.Phys.Chem.Lett.11.7}
C.~A. DelPo, B.~Kudisch, K.~H. Park, S.-U.-Z. Khan, F.~Fassioli, D.~Fausti,
  B.~P. Rand, and G.~D. Scholes, \enquote{Polariton transitions in femtosecond
  transient absorption studies of ultrastrong lightšcmolecule coupling,}
  {\protect\JournalTitle{J. Phys. Chem. Lett.}} \textbf{11}, 2667--2674 (2020).

\bibitem{PhysRevLett.123.247701}
G.~A. Peterson, S.~Kotler, F.~Lecocq, K.~Cicak, X.~Y. Jin, R.~W. Simmonds,
  J.~Aumentado, and J.~D. Teufel, \enquote{Ultrastrong parametric coupling
  between a superconducting cavity and a mechanical resonator,}
  {\protect\JournalTitle{Phys. Rev. Lett.}} \textbf{123}, 247701 (2019).

\bibitem{s41467-020-16524-x}
D.~G. Baranov, B.~Munkhbat, E.~Zhukova, A.~Bisht, A.~Canales, B.~Rousseaux,
  G.~Johansson, T.~J. Antosiewicz, and T.~Shegai, \enquote{Ultrastrong coupling
  between nanoparticle plasmons and cavity photons at ambient conditions,}
  {\protect\JournalTitle{Nature Communications}} \textbf{11}, 2715 (2020).

\bibitem{PhysRevA.55.3117}
K.~Banaszek and K.~W{\'o}dkiewicz, \enquote{Operational theory of homodyne
  detection,} {\protect\JournalTitle{Physical Review A}} \textbf{55}, 3117
  (1997).

\bibitem{Li:22}
M.~Li, H.~Zhang, J.~Lu, and Z.~hua Shen, \enquote{Laser ultrasonic loading and
  optical interference detection of closed cracks in k9 glass,}
  {\protect\JournalTitle{Opt. Lett.}} \textbf{47}, 3736--3739 (2022).

\bibitem{Nanoscale2022.14}
R.~C. Ng, A.~El~Sachat, F.~Cespedes, M.~Poblet, G.~Madiot,
  J.~Jaramillo-Fernandez, O.~Florez, P.~Xiao, M.~Sledzinska, C.~M.
  Sotomayor-Torres, and E.~Chavez-Angel, \enquote{Excitation and detection of
  acoustic phonons in nanoscale systems,} {\protect\JournalTitle{Nanoscale}}
  \textbf{14}, 13428--13451 (2022).

\bibitem{Nat.Commun.12.6206}
M.~Barra-Burillo, U.~Muniain, S.~Catalano, M.~Autore, F.~Casanova, L.~E. Hueso,
  J.~Aizpurua, R.~Esteban, and R.~Hillenbrand, \enquote{Microcavity phonon
  polaritons from the weak to the ultrastrong phonon–photon coupling regime,}
  {\protect\JournalTitle{Nat. Commun.}} \textbf{12}, 6206 (2021).

\bibitem{Comp.Phys.Comm.184(2013)1234}
J.~R. Johansson, P.~D. Nation, and F.~Nori, \enquote{Qutip: An open-source
  python framework for the dynamics of open quantum systems,}
  {\protect\JournalTitle{Comp. Phys. Comm.}} \textbf{183}, 1760--1772 (2013).

\end{thebibliography}

\end{document}